\documentclass[twocolumn,showpacs,preprintnumbers,amsmath,amssymb]{revtex4}

\usepackage{graphicx}
\usepackage{dcolumn}
\usepackage{bm}

\usepackage{amssymb}
\usepackage{amsmath}

\begin{document}


\title{Horizon thermodynamics in holographic cosmological models with a power-law term}

\author{Nobuyoshi {\sc Komatsu}}  \altaffiliation{E-mail: komatsu@se.kanazawa-u.ac.jp} 
\affiliation{Department of Mechanical Systems Engineering, Kanazawa University, Kakuma-machi, Kanazawa, Ishikawa 920-1192, Japan}

\date{\today}

\begin{abstract}

Thermodynamics on the horizon of a flat universe at late times is studied in holographic cosmological models that assume an associated entropy on the horizon.
In such models, a $\Lambda(t)$ model similar to a time-varying $\Lambda(t)$ cosmology is favored because of the consistency of energy flows across the horizon. 
Based on this consistency, a $\Lambda(t)$ model with a power-law term proportional to $H^{\alpha}$ is formulated to systematically examine the evolution of the Bekenstein--Hawking entropy.
Here, $H$ is the Hubble parameter and $\alpha$ is a free parameter whose value is a real number.
The present model always satisfies the second law of thermodynamics on the horizon.
In particular, the universe for $\alpha <2$ tends to approach thermodynamic equilibrium-like states.
Consequently, when $\alpha < 2$, the maximization of the entropy should be satisfied as well, at least in the last stage of the evolution of an expanding universe.
A relaxation-like process before the last stage is also examined from a thermodynamics viewpoint.

\end{abstract}


\maketitle

\section{Introduction} 
\label{Introduction}

The $\Lambda$CDM (Lambda cold dark matter) model can elegantly explain the accelerated expansion of the late universe \cite{PERL1998_Riess1998,Planck2015,Riess2016,Planck2018}.
In this model, an extra driving term, i.e., a cosmological constant $\Lambda$ related to dark energy, is added to the Friedmann and acceleration equations. 
However, the $\Lambda$CDM model suffers from several difficulties \cite{Weinberg1989}.
To resolve these difficulties, various models have been proposed \cite{Bamba1}: e.g.,
$\Lambda (t)$CDM models (i.e., a time-varying $\Lambda(t)$ cosmology) \cite{Freese-Mimoso_2015,Sola_2009-2015,Nojiri2006,Sola_2015L14,Valent2015,Sola_2017-2018,Sola2019}, 
bulk viscous models \cite{Weinberg0,Barrow11-Zimdahl1,Brevik1-Nojiri1,Barrow21,Avelino2-Brevik}, and creation of CDM (CCDM) models \cite{Prigogine_1988-1989,Lima1992-1996,LimaOthers2001-2016}. 
In addition, thermodynamic scenarios based on the holographic principle \cite{Hooft-Bousso} have been recently proposed \cite{Easson,Cai,Basilakos1,Basilakos2014-Gohar,Gohar_a_b,Koma45,Koma6,Koma78,Koma9,Sheykhi1,Sadjadi1,Padma2012AB,Cai2012-Tu2013,Tu2013-2015,Hadi-Sheykhia2018,Neto2018a,Koma10,Koma11,Koma12,Sheykhi2,Karami2011-2018,Padmanabhan2004,ShuGong2011}. 
Extending the concept of black hole thermodynamics, these scenarios assume that the horizon of the universe has an associated entropy:
e.g., the Bekenstein--Hawking entropy \cite{Bekenstein1Hawking1}, the Tsallis--Cirto entropy \cite{Tsallis2012}, a modified R\'{e}nyi entropy \cite{Czinner1,Czinner2}, and power-law corrections \cite{Das2008,Radicella2010}.

These entropies have been applied to holographic cosmological models and have led to extra driving terms expressed as a function of the Hubble parameter $H$.
For example, in entropic cosmology \cite{Easson}, $H^{2}$ and $H$ terms are derived using the Bekenstein--Hawking entropy and the Tsallis--Cirto entropy, respectively \cite{Koma9}.
In addition, a power-law term proportional to $H^{\alpha}$ can be derived from Padmanabhan's holographic equipartition law with a power-law corrected entropy \cite{Koma11}.
Cosmological models with $H^{\alpha}$ terms are expected to be suitable for systematic studies because $\alpha$ can be a free parameter whose value is a real number.

Formulations of such models are generally categorized into two types \cite{Koma6}.
The first type is $\Lambda(t)$, which is similar to $\Lambda(t)$CDM models \cite{Freese-Mimoso_2015,Sola_2009-2015,Nojiri2006,Sola_2015L14,Valent2015,Sola_2017-2018,Sola2019}. 
In $\Lambda(t)$ models, both the Friedmann equation and the acceleration equation include an extra driving term \cite{Koma6}.
The second type is BV, which is similar to both bulk viscous models \cite{Weinberg0,Barrow11-Zimdahl1,Brevik1-Nojiri1,Barrow21,Avelino2-Brevik} and CCDM models \cite{Prigogine_1988-1989,Lima1992-1996,LimaOthers2001-2016}. In BV models, the acceleration equation includes an extra driving term whereas the Friedmann equation does not \cite{Koma6}.
For example, Basilakos \textit{et al.} have shown that the $H^{2}$ terms in $\Lambda(t)$ models do not describe structure formations properly \cite{Basilakos1}.
(A power series of $H$ has been examined in $\Lambda(t)$ models; see, e.g., the works of Sol\`{a} \textit{et al.} \cite{Sola_2015L14}, G\'{o}mez-Valent \textit{et al.} \cite{Valent2015}, and Rezaei \textit{et al.} \cite{Sola2019}.)
Li and Barrow have reported that the $H$ terms in BV models are difficult to reconcile with astronomical observations of structure formations \cite{Barrow21}.
The present author and Kimura have indicated that $\Lambda(t)$ models are consistent with the observed growth rate of clustering, unlike BV models \cite{Koma6}.
Observations thus imply that $\Lambda(t)$ models are suitable for holographic cosmological models.
However, the suitability of models  has not been discussed from a theoretical viewpoint.

The evolution of the universe is expected to be related to thermodynamics on the horizon.
It is well known that ordinary, isolated macroscopic systems spontaneously evolve to equilibrium states of the maximum entropy consistent with their constraints \cite{Callen}.
In other words, the entropy of such systems does not decrease (i.e., the second law of thermodynamics) and approaches a certain maximum value in the last stage (i.e., the maximization of entropy).
The second law of thermodynamics has been extensively studied from a cosmological viewpoint \cite{Koma11,Koma12,Sheykhi1,Sadjadi1,Sheykhi2,Karami2011-2018,Padmanabhan2004,ShuGong2011,Easther1,Barrow3,Davies11_Davis0100,Gong00_01,Egan1}.
The maximization of entropy has recently attracted attention \cite{Pavon2013,Mimoso2013,Krishna2017,Krishna2019,Bamba2018,Pavon2019,Saridakis2019,deSitter}.  
For example, Mimoso and Pav\'{o}n have examined the maximization of entropy in the universe with a de Sitter era and have shown that the universe behaves as an ordinary macroscopic system at least in the last stage \cite{Mimoso2013}.
Krishna and Mathew have investigated entropy maximization and the holographic equipartition law \cite{Krishna2017,Krishna2019}.
Bamba \textit{et al.} have examined thermodynamic equilibrium conditions for several entropies in Rastall gravity \cite{Bamba2018}.
The previous works imply that a certain type of universe behaves as an ordinary macroscopic system in the last stage \cite{Mimoso2013}.
Accordingly, the evolution of the universe should be a kind of relaxation process.
Such a relaxation-like process has not yet been examined systematically from a thermodynamics viewpoint.

In this context, we study thermodynamics on the horizon of the universe in a holographic cosmological model that includes a power-law term.
To examine the suitability of models, we derive energy flows across the horizon using two methods.
Based on the consistency of the two energy flows, we formulate a cosmological model with the power-law term.
Using this model, we systematically examine not only the background evolution of the late universe but also thermodynamics on the horizon.
The present study should facilitate the discussion of various cosmological models from a thermodynamics viewpoint.
Note that density perturbations related to structure formations are not discussed here.

The remainder of the present article is organized as follows.
In Sec.\ \ref{General cosmological equations and energy flow across the horizon}, a general formulation of cosmological equations in a flat Friedmann--Robertson--Walker (FRW) universe is reviewed.
An energy flow across the horizon is derived from the general formulation.
In Sec.\ \ref{Thermodynamics on the horizon}, the Bekenstein--Hawking entropy is reviewed.
A similar energy flow is derived from the equipartition law of energy and the consistency of the two derived energy flows is discussed.
In Sec.\ \ref{Power-law term}, a cosmological model that includes a power-law term is formulated based on the consistency.
The background evolution of the late universe in the present model is examined.
In Sec.\ \ref{SL}, the entropy evolution for the present model is examined to discuss the second law of thermodynamics and the maximization of entropy. 
Finally, in Sec.\ \ref{Conclusions}, the conclusions of the study are presented.


\section{General cosmological equations and energy flow across the horizon} 
\label{General cosmological equations and energy flow across the horizon}

We consider a homogeneous, isotropic, and spatially flat universe, i.e., a flat FRW universe.
In Sec.\ \ref{General cosmological equations}, we review a general formulation of the cosmological equations \cite{Koma9} because it can be used for holographic cosmological models.
Based on the general formulation, an energy flow across the horizon is discussed in Sec.\ \ref{Energy flow across the Hubble horizon}.

\subsection{General Friedmann, acceleration, and continuity equations in a flat FRW universe} 
\label{General cosmological equations}

We review a general formulation of the cosmological equations in a flat FRW universe using the scale factor $a(t)$ at time $t$.
According to Ref.\ \cite{Koma9}, the general Friedmann equation is given as 
\begin{equation}
 H(t)^2      =  \frac{ 8\pi G }{ 3 } \rho (t)    + f_{\Lambda}(t)            ,                                                 
\label{eq:General_FRW01_f_0} 
\end{equation} 
and the general acceleration equation is 
\begin{align}
  \frac{ \ddot{a}(t) }{ a(t) }   &=  \dot{H}(t) + H(t)^{2}                                                                        \notag \\
                                          &=  -  \frac{ 4\pi G }{ 3 } \left ( \rho (t) + \frac{3p(t)}{c^2}  \right ) +   f_{\Lambda}(t)    +  h_{\textrm{B}}(t)   \notag \\
                                          &=  -  \frac{ 4\pi G }{ 3 }  ( 1+  3w ) \rho (t)                                   +   f_{\Lambda}(t)    +  h_{\textrm{B}}(t)  ,  
\label{eq:General_FRW02_g_0}
\end{align}
where the Hubble parameter $H(t)$ is defined by
\begin{equation}
   H(t) \equiv   \frac{ da/dt }{a(t)} =   \frac{ \dot{a}(t) } {a(t)}  , 
\label{eq:Hubble}
\end{equation}
and $w$ represents the equation of the state parameter for a generic component of matter, which is given as  
\begin{equation}
  w = \frac{ p(t) } { \rho(t)  c^2 }    .
\label{eq:w}
\end{equation}
Here, $G$, $c$, $\rho(t)$, and $p(t)$ are the gravitational constant, the speed of light, the mass density of cosmological fluids, and the pressure of cosmological fluids, respectively \cite{Koma9}.
For a matter-dominated universe and a radiation-dominated universe, $w$ is $0$ and $1/3$, respectively.
We consider the matter-dominated universe, i.e., $w =0$, although $w$ is retained for generality.
Two extra driving terms, i.e., $f_{\Lambda}(t)$ and $h_{\textrm{B}}(t)$, are phenomenologically assumed.
In this study, $f_{\Lambda}(t)$ is used for the $\Lambda (t)$ model,  similar to $\Lambda(t)$CDM models, whereas $h_{\textrm{B}}(t)$ is used for the BV model, similar to bulk viscous models and CCDM models.
Accordingly, we set $h_{\textrm{B}}(t) =0$ for the $\Lambda (t)$ model and $f_{\Lambda}(t) = 0$ for the BV model.

Coupling [$(1+3w) \times $ Eq.\ (\ref{eq:General_FRW01_f_0})] with [$2 \times $ Eq.\ (\ref{eq:General_FRW02_g_0})] and rearranging the resultant equation, we obtain the differential equation given by
\begin{equation}
    \dot{H} = - \frac{3}{2} (1+w) H^{2}  +  \frac{3}{2} (1+w)    f_{\Lambda}(t)     + h_{\textrm{B}}(t)   .  
\label{eq:Back2}
\end{equation}
Using this equation, we can examine the background evolution of the universe in various cosmological models.

\begin{figure} [t] 
\begin{minipage}{0.495\textwidth}
\begin{center}
\scalebox{0.33}{\includegraphics{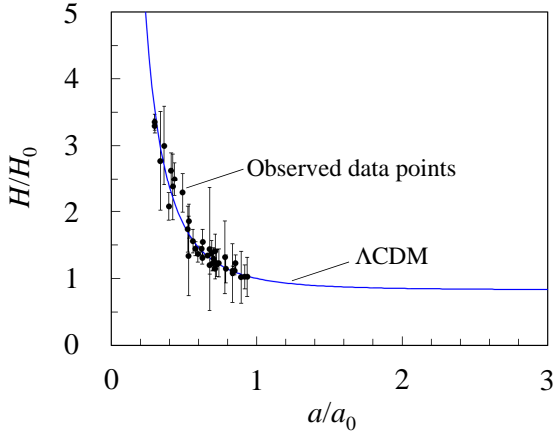}}
\end{center}
\end{minipage}
\caption{ (Color online). Evolution of the normalized Hubble parameter $H/H_{0}$ for a fine-tuned $\Lambda$CDM model.
The horizontal axis is the normalized scale factor $a/a_{0}$.
The solid line represents the $\Lambda$CDM model given by Eq.\ (\ref{eq:Sol_H_LCDM_1}), in which $\Omega_{\Lambda}$ is set to $0.685$ from the Planck 2018 results \cite{Planck2018}.
The closed circles with error bars are observed data points taken from Ref.\ \cite{Hubble2017}. 
The data points are normalized as $H/H_{0}$, where $H_{0}$ is set to $67.4$ km/s/Mpc from Ref.\ \cite{Planck2018}. }
\label{Fig-H-a_1}
\end{figure}

For example, for $\Lambda$CDM models, substituting $f_{\Lambda} (t) = \Lambda / 3$, $ h_{\textrm{B}} (t) =0$, and $w=0$ into Eq.\ (\ref{eq:Back2}) yields 
\begin{equation}
    \dot{H} = - \frac{3}{2} H^{2}  +  \frac{\Lambda}{2}    . 
\label{eq:Back2_LCDM}
\end{equation}
The evolution of $H$ is given by 
\begin{equation}
 \left (  \frac{H}{H_{0}} \right )^{2}  =   (1- \Omega_{\Lambda} )   \left ( \frac{a}{a_{0}} \right )^{ - 3}  + \Omega_{\Lambda}    ,
\label{eq:Sol_H_LCDM_1}
\end{equation}
where $\Omega_{\Lambda} (= \Lambda /( 3 H_{0}^{2} ) )$ is the density parameter for $\Lambda$. 
In a flat FRW universe, the density parameter for matter is given by $1- \Omega_{\Lambda}$, neglecting the influence of radiation.
$H_{0}$ and $a_{0}$ represent the Hubble parameter and the scale factor at the present time, respectively.
To observe the background evolution of the universe, the normalized Hubble parameter $H/H_{0}$ given by Eq.\ (\ref{eq:Sol_H_LCDM_1}) is shown in Fig.\ \ref{Fig-H-a_1}.
In this figure, $\Omega_{\Lambda}$ is set to a fine-tuned value, $0.685$, from the Planck 2018 results \cite{Planck2018}.
The observed data points \cite{Hubble2017} are also plotted. 
In an expanding universe, $a/a_{0}$ increases with time.
Note that we do not discuss the significant tension between the Planck results \cite{Planck2015} and the local (distance ladder) measurement from the Hubble Space Telescope \cite{Riess2016}.

As shown in Fig.\ \ref{Fig-H-a_1}, $H/H_{0}$ decreases with $a/a_{0}$ and gradually approaches a constant value.
The constant value is given by $\Omega_{\Lambda}^{1/2}$, which is calculated from Eq.\ (\ref{eq:Sol_H_LCDM_1}) by applying $ a/a_{0} \rightarrow \infty$.
These results imply that $\dot{H}$ is negative and gradually approaches zero.
That is, $\dot{H} < 0$ is satisfied and $\dot{H} \rightarrow 0$ is expected at least in the last stage. 
The last stage is similar to that in a de Sitter universe and should be in equilibrium-like states.
The thermodynamics of the de Sitter universe was examined in the work of Mimoso and Pav\'{o}n \cite{Mimoso2013}.
For similar discussions, see, e.g., Refs.\ \cite{Pavon2013,Krishna2017,Krishna2019,Bamba2018,Pavon2019,Saridakis2019,deSitter}.

In this way, the background evolution of the universe can be examined using Eq.\ (\ref{eq:Back2}) derived from the general Friedmann and acceleration equations.
In addition, from the two equations, we can calculate the general continuity equation written as \cite{Koma9} 
\begin{equation}
       \dot{\rho} + 3  H (1+w)  \rho   =  \frac{3}{4 \pi G} H \left(  h_{\textrm{B}}(t)  -  \frac{\dot{f}_{\Lambda}(t)}{2 H }      \right )                   .
\label{eq:drho_General_00}
\end{equation}
The right-hand side of this equation is nonzero, except for $\Lambda$CDM models.
In $\Lambda (t)$ models, such as $\Lambda (t)$CDM models, the nonzero right-hand side can be interpreted as a kind of energy exchange cosmology \cite{Barrow22}.
In BV models, such as CCDM models, the nonzero right-hand side is related to an effective pressure \cite{Prigogine_1988-1989,Lima1992-1996,LimaOthers2001-2016}.

\subsection{Energy flow across the Hubble horizon derived from the general formulation} 
\label{Energy flow across the Hubble horizon}

In this subsection, we examine an energy flow across the horizon using the general formulation according to previous studies \cite{Easson,ShuGong2011,Mimoso2018}.
To this end, we consider an energy flow across a spherical surface of Hubble horizon (radius) $r_{H}$, which is given by 
\begin{equation}
     r_{H} = \frac{c}{H}   .
\label{eq:rH}
\end{equation}
In the present study, the Hubble horizon is equivalent to an apparent horizon because a flat FRW universe is considered \cite{Easson}.

When the universe expands at the Hubble rate, the energy flow across the horizon \cite{Easson,ShuGong2011,Mimoso2018} can be written as 
\begin{align}
-\dot{E} = - \frac{dE}{dt}  &= A_{H} \left ( \rho  + \frac{p}{c^{2}} \right ) c^{2} H r_{H}      \notag \\
                                     &= A_{H} r_{H} c^{2} (1+w) \rho H    ,
\label{eq:-dEdt}
\end{align}
where $A_{H}$ is the surface area of the sphere with the Hubble horizon $r_{H}$, and  $w = p / (\rho c^2 )$ is given by Eq.\ (\ref{eq:w}).
The right-hand side of this equation can be calculated using Eqs.\ (\ref{eq:General_FRW01_f_0}), (\ref{eq:drho_General_00}), and (\ref{eq:rH}) and $A_{H}=4 \pi r_{H}^{2}$.
The detailed calculation is given in Appendix\ \ref{Solution-dEdt}.
From Eqs.\ (\ref{eq:-dEdt2_A}) and (\ref{eq:-dEdt3_A}), the energy flow is written as 
\begin{align}
-\dot{E}     
&=  \frac{ A_{H} c^{3} }{4 \pi G }  \left (  -  \dot{H}    +  h_{\textrm{B}}(t)  \right )    \notag \\
                &=  \frac{ c^{5} }{ G }  \left (  -  \frac{ \dot{H} } {H^{2}}    +  \frac{ h_{\textrm{B}}(t) }{H^{2}}  \right )    .
\label{eq:-dEdt_Sol}
\end{align}
This equation indicates that $-\dot{E}$ does not depend on $f_{\Lambda} (t) $ for $\Lambda (t)$ models.
That is, $-\dot{E}$ is proportional to $ - \dot{H}/H^{2}$ for $\Lambda (t)$ models because $h_{\textrm{B}}(t) =0$.
Therefore, $-\dot{E}$ is expected to approach zero if $H$ gradually approaches a positive constant, as shown in Fig.\ \ref{Fig-H-a_1}.
In the next section, we examine this expectation.
We also derive a similar energy flow using another method and discuss the consistency of the two derived energy flows.

\section{Thermodynamics on the horizon} 
\label{Thermodynamics on the horizon}

Based on the holographic principle \cite{Hooft-Bousso}, we assume that the horizon of the universe has an associated entropy and an approximate temperature \cite{Easson}.
In Sec.\ \ref{Bekenstein-Hawking entropy}, the Bekenstein--Hawking entropy is introduced.
In Sec.\ \ref{Energy flow Equipartition}, an energy flow across the horizon is derived assuming the equipartition law of energy.

Note that the entropy $S$ of ordinary isolated macroscopic systems does not decrease, i.e., $\dot{S} \ge 0$, as mentioned previously.
In particular, the entropy should approach a certain maximum value consistent with their constraints, i.e., $\ddot{S} < 0$ \cite{Callen}.
A certain type of universe, e.g., a de Sitter universe \cite{Pavon2013,Mimoso2013,Krishna2017,Krishna2019,deSitter}, behaves as an ordinary macroscopic system \cite{Mimoso2013}.
For related studies, see, e.g., the works of Pav\'{o}n and Radicella \cite{Pavon2013}, Mimoso and Pav\'{o}n \cite{Mimoso2013}, and Krishna and Mathew \cite{Krishna2017,Krishna2019}.

\subsection{Entropy on the Hubble horizon} 
\label{Bekenstein-Hawking entropy}

In this study, the Bekenstein--Hawking entropy $S_{\rm{BH}}$ \cite{Bekenstein1Hawking1} is applied to an associated entropy on the Hubble horizon because it is the most standard one.
At late times, $S_{\rm{BH}}$ is dozens of orders of magnitude larger than other entropies related to matter, radiation, etc., as examined by Egan and Lineweaver \cite{Egan1}.
That is, $S_{\rm{BH}}$ should be approximately equivalent to the total entropy of the late universe. 
Accordingly, we focus on $S_{\rm{BH}}$ and do not discuss the generalized second law of thermodynamics.
Note that the Hubble horizon is equivalent to an apparent horizon because a flat universe is considered.

The Bekenstein--Hawking entropy is written as
\begin{equation}
 S_{\rm{BH}}  = \frac{ k_{B} c^3 }{  \hbar G }  \frac{A_{H}}{4}   ,
\label{eq:SBH}
\end{equation}
where $k_{B}$ and $\hbar$ are the Boltzmann constant and the reduced Planck constant, respectively. 
The reduced Planck constant is defined as $\hbar \equiv h/(2 \pi)$, where $h$ is the Planck constant \cite{Koma10,Koma11,Koma12}.
Substituting $A_{H}=4 \pi r_{H}^2 $ into Eq.\ (\ref{eq:SBH}) and applying Eq.\ (\ref{eq:rH}) yields
\begin{equation}
S_{\rm{BH}}  = \frac{ k_{B} c^3 }{  \hbar G }   \frac{A_{H}}{4}       
                  =  \left ( \frac{ \pi k_{B} c^5 }{ \hbar G } \right )  \frac{1}{H^2}  
                  =    \frac{K}{H^2}    , 
\label{eq:SBH2}      
\end{equation}
where $K$ is a positive constant given by
\begin{equation}
  K =  \frac{  \pi  k_{B}  c^5 }{ \hbar G } = \frac{  \pi  k_{B}  c^2 }{ L_{p}^{2} }  , 
\label{eq:K-def}
\end{equation}
and $L_{p}$ is the Planck length, which is written as \cite{Koma10,Koma11,Koma12}
\begin{equation}
  L_{p} = \sqrt{ \frac{\hbar G} { c^{3} } } .
\label{eq:Lp}
\end{equation}
From Eq.\ (\ref{eq:SBH2}), we can confirm $S_{\rm{BH}} >0$.

We now calculate the time derivative of $S_{\rm{BH}}$.
Differentiating Eq.\ (\ref{eq:SBH2}) with respect to $t$ yields the first derivative of $S_{\rm{BH}}$, which is given by \cite{Koma11,Koma12}
\begin{equation}
\dot{S}_{\rm{BH}}  =  \frac{d }{dt} S_{\rm{BH}}  = \frac{d}{dt}   \left ( \frac{K}{H^{2}} \right )  =  \frac{-2K \dot{H} }{H^{3}}                  ,
\label{eq:dSBH}      
\end{equation}
or equivalently
\begin{equation}
\dot{S}_{\rm{BH}}      = 2 S_{\rm{BH}} \left ( \frac{- \dot{H} }{H} \right ) .  
\label{eq:dSBH_2}      
\end{equation}
In addition, differentiating Eq.\ (\ref{eq:dSBH}) with respect to $t$ yields the second derivative of $S_{\rm{BH}}$, which is given by
\begin{align}
\ddot{S}_{\rm{BH}}   &= \frac{d}{dt} \dot{S}_{\rm{BH}} = \frac{d}{dt} \left ( \frac{-2K \dot{H} }{H^{3}} \right )    
                               =  - 2 K  \left ( \frac{\ddot{H} }{H^{3}} - \frac{3 \dot{H}^{2} }{H^{4}}  \right )                  \notag  \\
                             &=    2 \frac{K}{H^{2}}  \left ( \frac{   3 \dot{H}^{2}  - \ddot{H} H   }{H^{2}} \right )     
                               =    2 S_{\rm{BH}}       \left ( \frac{   3 \dot{H}^{2}   - \ddot{H} H  }{H^{2}}  \right )     .
\label{eq:d2SB_1}      
\end{align}
Equation\ (\ref{eq:dSBH_2}) indicates that the sign of $\dot{S}_{\rm{BH}} $ depends on whether $- \dot{H} /H $ is positive or negative because $S_{\rm{BH}} > 0$.
Similarly, Eq.\ (\ref{eq:d2SB_1}) indicates that the sign of $\ddot{S}_{\rm{BH}}$ depends on whether $3 \dot{H}^{2}   - \ddot{H} H$ is positive or negative.  
Accordingly, the two signs depend on the evolution of $H$ in cosmological models.

\begin{figure} [b] 
\begin{minipage}{0.495\textwidth}
\begin{center}
\scalebox{0.34}{\includegraphics{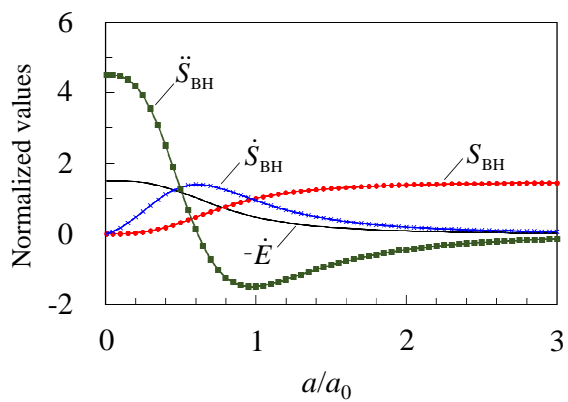}}
\end{center}
\end{minipage}
\caption{ (Color online). Evolution of $S_{\rm{BH}}$, $\dot{S}_{\rm{BH}}$, $\ddot{S}_{\rm{BH}}$, and $- \dot{E}$ for a fine-tuned $\Lambda$CDM model for $\Omega_{\Lambda} =0.685$.
The four parameters are normalized as $S_{\rm{BH}} (H_{0}^{2}/K)$, $\dot{S}_{\rm{BH}} (H_{0}/K)$, $\ddot{S}_{\rm{BH}} (1/K)$, and $- \dot{E} (G/c^{5})$.}
\label{Fig-S_dSdt_dEdt-a_LCDM}
\end{figure}

We now show the evolution of the entropy on the horizon of the universe in a typical model.
To this end, we consider a fine-tuned $\Lambda$CDM model given by Eqs.\ (\ref{eq:Back2_LCDM}) and (\ref{eq:Sol_H_LCDM_1}).
As shown in Fig.\ \ref{Fig-H-a_1}, $\Omega_{\Lambda}$ is set to $0.685$ and the influence of radiation is neglected.
The evolution of $S_{\rm{BH}}$, $\dot{S}_{\rm{BH}}$, and $\ddot{S}_{\rm{BH}}$ is plotted in Fig.\ \ref{Fig-S_dSdt_dEdt-a_LCDM}.
As shown in this figure, $S_{\rm{BH}}$ increases with $a/a_{0}$.
However, when $a/a_{0} \gtrapprox  0.6$, the increase in $S_{\rm{BH}}$ tends to become gradually slower.
These results depend on the evolution of $H$.
(As examined in Fig.\ \ref{Fig-H-a_1}, $H/H_{0}$ decreases with $a/a_{0}$ and gradually approaches a constant value.
That is, we have $\dot{H} < 0$ and expect $\dot{H} \rightarrow 0$ in the last stage.)

In addition, as shown in Fig.\ \ref{Fig-S_dSdt_dEdt-a_LCDM}, $\dot{S}_{\rm{BH}}$ is positive because $S_{\rm{BH}}$ increases with $a/a_{0}$.
Therefore, the second law of thermodynamics, $\dot{S}_{\rm{BH}} \geq 0$, is satisfied in the $\Lambda$CDM model.
In addition, $\dot{S}_{\rm{BH}}$ gradually decreases with $a/a_{0}$ and approaches zero.
Consequently, the maximization of entropy, $\ddot{S}_{\rm{BH}} < 0$, is satisfied in the last stage even though $\ddot{S}_{\rm{BH}}$ is positive in the early stage ($a/a_{0} \lessapprox 0.6$).
Similar discussions are given in Refs.\ \cite{Pavon2013,Mimoso2013,Bamba2018,Pavon2019,Saridakis2019,Krishna2017,Krishna2019,deSitter}.
In this way, we can study the evolution of the entropy on the horizon.
In Sec.\ \ref{SL}, we systematically examine the second law and entropy maximization using a cosmological model with a power-law term.

Finally, the energy flow $-\dot{E}$ derived from the general formulation is examined.
As shown in Fig.\ \ref{Fig-S_dSdt_dEdt-a_LCDM}, the energy flow decreases with $a/a_{0}$ and gradually approaches zero.
This is because $- \dot{E}$ is proportional to $- \dot{H} /H^{2}$ from Eq.\ (\ref{eq:-dEdt_Sol}) when $ h_{\textrm{B}}(t) =0$ is used for the $\Lambda$CDM model.
In the next subsection, a similar energy flow is derived using another method.

\subsection{Energy flow across the Hubble horizon derived from the equipartition law of energy}
\label{Energy flow Equipartition}

In Sec.\ \ref{Energy flow across the Hubble horizon}, an energy flow across the horizon was derived from the general formulation.
In this subsection, a similar energy flow is derived from the equipartition law of energy \cite{Padma2010,ShuGong2011}.
The consistency of the two energy flows is discussed here.
The relationship between the energy flow and a deceleration parameter is also examined.

We have assumed that the information of the bulk is stored on the horizon based on the holographic principle \cite{Hooft-Bousso}.  
In addition, we assume the equipartition law of energy on the horizon according to Refs.\ \cite{Padma2010,ShuGong2011}.
Consequently, the energy $\mathcal{E}$ on the Hubble horizon can be given by \cite{Padma2010,ShuGong2011}
\begin{equation}
 \mathcal{E} =  N  \times \frac{1}{2} k_{B} T     , 
\label{E_equip}
\end{equation}
where $N$ is the number of degrees of freedom on a spherical surface of Hubble radius $r_{H}$, which is written as
\begin{equation}
  N = \frac{4 S_{\rm{BH}} }{k_{B}}       .
\label{N_sur}
\end{equation}
The temperature $T$ on the Hubble horizon is given by 
\begin{equation}
 T = \frac{ \hbar H}{   2 \pi  k_{B}  }   .
\label{eq:T_1}
\end{equation}
Substituting Eq.\ (\ref{N_sur}) into Eq.\ (\ref{E_equip}) yields
\begin{equation}
 \mathcal{E} =  \left ( \frac{4 S_{\rm{BH}} }{k_{B}} \right )     \frac{1}{2} k_{B} T  =  2 S_{\rm{BH}}  T  . 
\label{E_ST}
\end{equation}
This equation, $ \mathcal{E} =2 S_{\rm{BH}}  T $, has been examined by Padmanabhan \cite{Padmanabhan2004,Padma2010}.

We now reformulate the energy given by Eq.\ (\ref{E_ST}) using the Hubble parameter.
Substituting Eqs.\ (\ref{eq:SBH2}) and (\ref{eq:T_1}) into Eq.\ (\ref{E_ST}) yields
\begin{equation}
 \mathcal{E}  =  2  \left ( \frac{ \pi k_{B} c^5 }{ \hbar G } \right )  \frac{1}{H^2}  \times \left ( \frac{ \hbar H}{   2 \pi  k_{B}  }    \right )  =    \frac{ c^{5} }{ G }  \left ( \frac{1}{H} \right )  .
\label{E_ST2}
\end{equation}
Differentiating the energy given by Eq.\ (\ref{E_ST2}) with respect to $t$ yields the following energy flow $\dot{\mathcal{E}}$: 
\begin{equation}
 \dot{\mathcal{E}}  =    \frac{ c^{5} }{ G }   \left (  -  \frac{ \dot{H} } {H^{2}}   \right )    .  
\label{eq:dEdt_ST2}
\end{equation}
This equation indicates that $\dot{\mathcal{E}}$ is proportional to $ - \dot{H}/H^{2}$.

Recall $-\dot{E}$ given by Eq.\ (\ref{eq:-dEdt_Sol}).
We select a $\Lambda(t)$ model because $h_{\textrm{B}}(t) =0$.
Substituting $h_{\textrm{B}}(t) = 0$ into Eq.\ (\ref{eq:-dEdt_Sol}) yields the following equation, which is equivalent to Eq.\ (\ref{eq:dEdt_ST2}):
\begin{equation}
 -\dot{E}     =   \dot{\mathcal{E}}  =    \frac{ c^{5} }{ G }   \left (  -  \frac{ \dot{H} } {H^{2}}   \right )  .
\label{eq:-dEdt3_dEdt_ST_0}
\end{equation}
In the $\Lambda(t)$ model, $-\dot{E}$ agrees with $\dot{\mathcal{E}}$ even though they were derived using different methods.
The consistency related to Eq.\ (\ref{eq:-dEdt3_dEdt_ST_0}) was described in the works of Krishna and Mathew \cite{Krishna2019} and Shu and Gong \cite{ShuGong2011}; BV models were not discussed.

This consistency may imply that $\Lambda(t)$ models are suitable for describing holographic cosmological models.
In the present study, we phenomenologically formulate a $\Lambda(t)$ model that includes a power-law term.
We discuss this in the next section.
Hereafter, we consider a $\Lambda(t)$ model and use $-\dot{E}$ because $-\dot{E} = \dot{\mathcal{E}}$.

In general, the temporal deceleration parameter $q$ is defined by
\begin{equation}
q \equiv  - \left ( \frac{\ddot{a} } {a H^{2}} \right )  , 
\label{eq:q_def}
\end{equation}
where positive (negative) $q$ represents deceleration (acceleration) \cite{Koma10}. 
Substituting $\ddot{a}/a = \dot{H} + H^{2}$ into Eq.\ (\ref{eq:q_def}) yields
\begin{equation}
q =   -  \frac{ \dot{H} } {H^{2}}   -1     .
\label{eq:q_2}
\end{equation}
Substituting Eq.\ (\ref{eq:-dEdt3_dEdt_ST_0}) for the $\Lambda(t)$ model into Eq.\ (\ref{eq:q_2}) yields the simple relation given by 
\begin{equation}
q =    -\dot{E} \left ( \frac{G}{ c^{5} }  \right )       -1     ,
\label{eq:q_dEdt}
\end{equation}
where $-\dot{E} ( G / c^{5} )$ represents the normalized energy flow.
When $-\dot{E} ( G / c^{5} ) < 1$, $q$ is negative.
In this way, $q$ can be evaluated from $-\dot{E}$.
We examine this in the next section.

It should be noted that we assume the following three, to study thermodynamics on the horizon of the universe in a holographic cosmological model.
Firstly, the equipartition law of energy can be applied to a relaxation-like process.
Usually, the equipartition law of energy is valid in equilibrium states. 
For example, a de Sitter universe is considered to be in equilibrium-like states because $H$ is constant.
In the present study, we assume that the equipartition law of energy can be applied to the relaxation-like process.
Secondly, the energy on the horizon can be given by Eq.\ (\ref{E_equip}).
In general, a factor included in this equation should depend on the Hamiltonian of the system.
In Ref.\ \cite{Padma2010}, the factor was assumed to be $1/2$, to derive a simple relation given by Eq.\ (\ref{E_ST}).
Similarly, in this paper, the factor is assumed to be $1/2$.
Thirdly, the temperature given by Eq.\ (\ref{eq:T_1}) can be applied to the relaxation-like process.
In fact, the original equation for the temperature was derived, using a de Sitter space in which $H$ is constant \cite{GibbonsHawking1977}.
Therefore, exactly speaking, Eq.\ (\ref{eq:T_1}) should be modified when the relaxation-like process is discussed.
The three assumptions have not yet been established although they were used in previous works.
In this study, the three assumptions are considered to be a viable scenario and are used for a $\Lambda(t)$ model.

\section{$\Lambda(t)$ model with a power-law term} 
\label{Power-law term}

The formulation of a $\Lambda(t)$ model is likely suitable for describing holographic cosmological models, as discussed in Sec.\ \ref{Energy flow Equipartition}.
Accordingly, in this study, we formulate a $\Lambda(t)$ model that includes a power-law term.
With $f_{\Lambda}(t)$ replaced by $f_{\alpha} (H)$, the Friedmann and acceleration equations for the present model are written as 
\begin{equation}
    H^2     =  \frac{ 8\pi G }{ 3 } \rho  +  f_{\alpha} (H)  ,  
\label{FRW01_power}
\end{equation}
and
\begin{equation}
  \frac{ \ddot{a} }{ a }           =   -  \frac{ 4 \pi G }{ 3} (1 + 3w)  \rho  +   f_{\alpha} (H)  .
\label{FRW02_power}
\end{equation}
From these two equations, the differential equation corresponding to Eq.\ (\ref{eq:Back2}) is 
\begin{equation}
    \dot{H} = - \frac{3}{2} (1+w) H^{2}  +  \frac{3}{2} (1+w)   f_{\alpha} (H)    , 
\label{eq:Back_power}
\end{equation}
where the extra driving term $f_{\alpha} (H)$ is given by \cite{Koma11}
\begin{equation}
        f_{\alpha} (H)  =   \Psi_{\alpha} H_{0}^{2} \left (  \frac{H}{H_{0}} \right )^{\alpha}  .
\label{eq:fa2}
\end{equation}
Here, $\alpha$ and $\Psi_{\alpha}$ are dimensionless constants whose values are real numbers.
The power-law term can be obtained from Padmanabhan's holographic equipartition law \cite{Padma2012AB} with a power-law corrected entropy \cite{Das2008}, as examined in a previous work \cite{Koma11}.
In the present paper, $\alpha$ and $\Psi_{\alpha}$ are considered to be independent free parameters.
That is, $\Psi_{\alpha}$ is assumed to be a kind of density parameter for effective dark energy.
In addition, $\Psi_{\alpha}$ is assumed to be 
\begin{equation}
       0 \leq \Psi_{\alpha} \leq 1 . 
\label{eq:psi0}
\end{equation}
The constant, $H$, and $H^{2}$ terms are obtained from Eq.\ (\ref{eq:fa2}) by applying $\alpha=0$, $1$, and $2$, respectively.
A power series of the Hubble parameter has been examined in $\Lambda(t)$ models \cite{Sola_2015L14,Valent2015,Sola2019}. 
It should be noted that in this study, $\alpha$ is a real number, unlike in previous works.

We now consider a matter-dominated universe.
Substituting $w =0$ and Eq.\ (\ref{eq:fa2}) into Eq.\ (\ref{eq:Back_power}) yields the differential equation for the present model given by 
\begin{align}
    \dot{H} &= - \frac{3}{2} H^{2}  +  \frac{3}{2}   \Psi_{\alpha} H_{0}^{2} \left (  \frac{H}{H_{0}} \right )^{\alpha}      \notag \\
               &= - \frac{3}{2} H^{2}  \left (  1 -   \Psi_{\alpha} \left (  \frac{H}{H_{0}} \right )^{\alpha -2} \right )      .  
\label{eq:Back_power_11}
\end{align}
The background evolution of the universe is calculated from Eq.\ (\ref{eq:Back_power_11}).
The solution method is summarized in Appendix\ \ref{Solution1}.
When $w=0$, from Eqs.\ (\ref{eq:Sol_HH0_power2}) and (\ref{eq:Sol_HH0_aa0_H2_A}), the solution for $\alpha \neq 2$ can be written as 
\begin{equation}  
    \left ( \frac{H}{H_{0}} \right )^{2-\alpha}  =   (1- \Psi_{\alpha})   \left ( \frac{a}{a_{0}} \right )^{ - \frac{3(2-\alpha)}{2}  }  + \Psi_{\alpha}   ,
\label{eq:Sol_HH0_power}
\end{equation}
and the solution for $\alpha = 2$ is
\begin{equation}
     \frac{H}{H_{0}}  =     \left ( \frac{a}{a_{0}} \right )^{ - \frac{3 (1- \Psi_{\alpha}) }{2}  }     .
\label{eq:Sol_HH0_aa0_H2}
\end{equation}
Note that Eq.\ (\ref{eq:Sol_HH0_power}) should reduce to Eq.\ (\ref{eq:Sol_HH0_aa0_H2}) when $\alpha \rightarrow 2$ is applied to Eq.\ (\ref{eq:Sol_HH0_power}).

We can examine various models using the above solutions.
For example, $\Lambda$CDM models are obtained from Eq.\ (\ref{eq:Sol_HH0_power}).
In this case, $ f_{\alpha} (H)$ is $\Lambda / 3$.
Substituting $\alpha =0$ into Eq.\ (\ref{eq:Sol_HH0_power}) and replacing $\Psi_{\alpha}$ by $\Omega_{\Lambda}$ yields
\begin{equation}
 \left (  \frac{H}{H_{0}} \right )^{2}  =   (1- \Omega_{\Lambda} )   \left ( \frac{a}{a_{0}} \right )^{ - 3}  + \Omega_{\Lambda}    ,
\label{eq:Sol_H_LCDM}
\end{equation}
where $\Omega_{\Lambda}$ is given by $\Lambda /( 3 H_{0}^{2} ) $. 
This equation is equivalent to the solution given by Eq.\ (\ref{eq:Sol_H_LCDM_1}).

We calculate the energy flow $ -\dot{E} $ across the horizon and a temporal deceleration parameter $q$ in the present model.
Recall that $ -\dot{E} $ given by Eq.\ (\ref{eq:-dEdt3_dEdt_ST_0}) and $q$ given by Eq.\ (\ref{eq:q_2}) both include $-\dot{H}/H^{2}$ terms.
This term is  directly obtained from Eq.\ (\ref{eq:Back_power_11}).
Substituting Eq.\ (\ref{eq:Back_power_11}) into Eq.\ (\ref{eq:-dEdt3_dEdt_ST_0}) yields the following energy flow:
\begin{equation}
 -\dot{E}     =   \frac{ c^{5} }{ G }   \left (  -  \frac{ \dot{H} } {H^{2}}   \right )   =  \frac{3}{2}   \frac{ c^{5} }{ G }   \left (  1 -   \Psi_{\alpha} \left (  \frac{H}{H_{0}} \right )^{\alpha -2} \right )    .
\label{eq:-dEdt3_dEdt_ST_0_power}
\end{equation}
In addition, substituting Eq.\ (\ref{eq:-dEdt3_dEdt_ST_0_power}) into Eq.\ (\ref{eq:q_dEdt}) yields
\begin{align}
q  &= -\dot{E} \left ( \frac{G}{ c^{5} }  \right )       -1                  =  \frac{3}{2}    \left (  1 -   \Psi_{\alpha} \left (  \frac{H}{H_{0}} \right )^{\alpha -2} \right )             -1     \notag \\
    &=  \frac{1}{2}  -   \frac{3}{2}     \Psi_{\alpha} \left (  \frac{H}{H_{0}} \right )^{\alpha -2}       .
\label{eq:q_power}
\end{align}

\begin{figure}[b] 
\begin{minipage}{0.495\textwidth}
\begin{center}
\scalebox{0.34}{\includegraphics{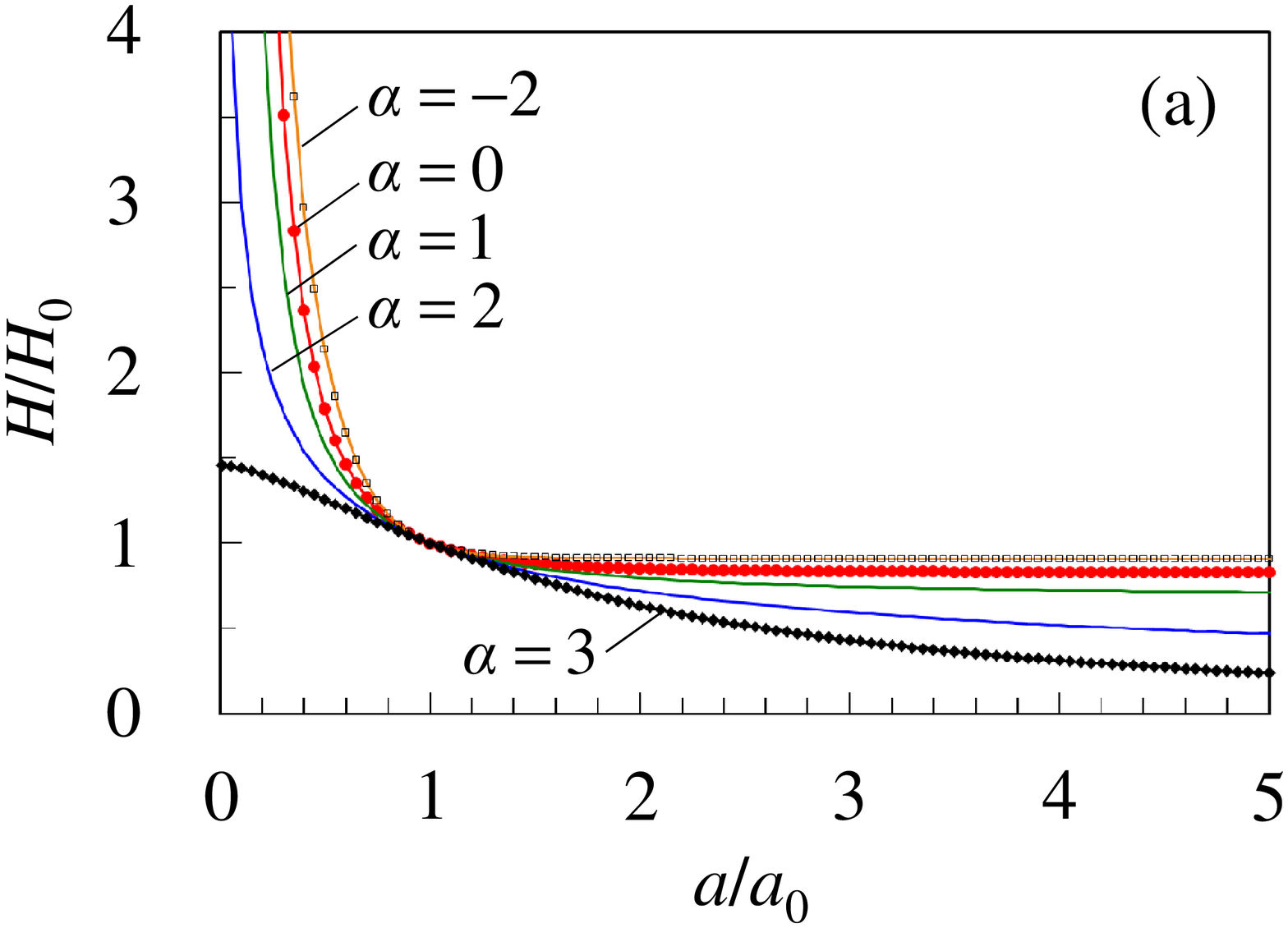}}
\end{center}
\end{minipage}
\begin{minipage}{0.495\textwidth}
\begin{center}
\scalebox{0.34}{\includegraphics{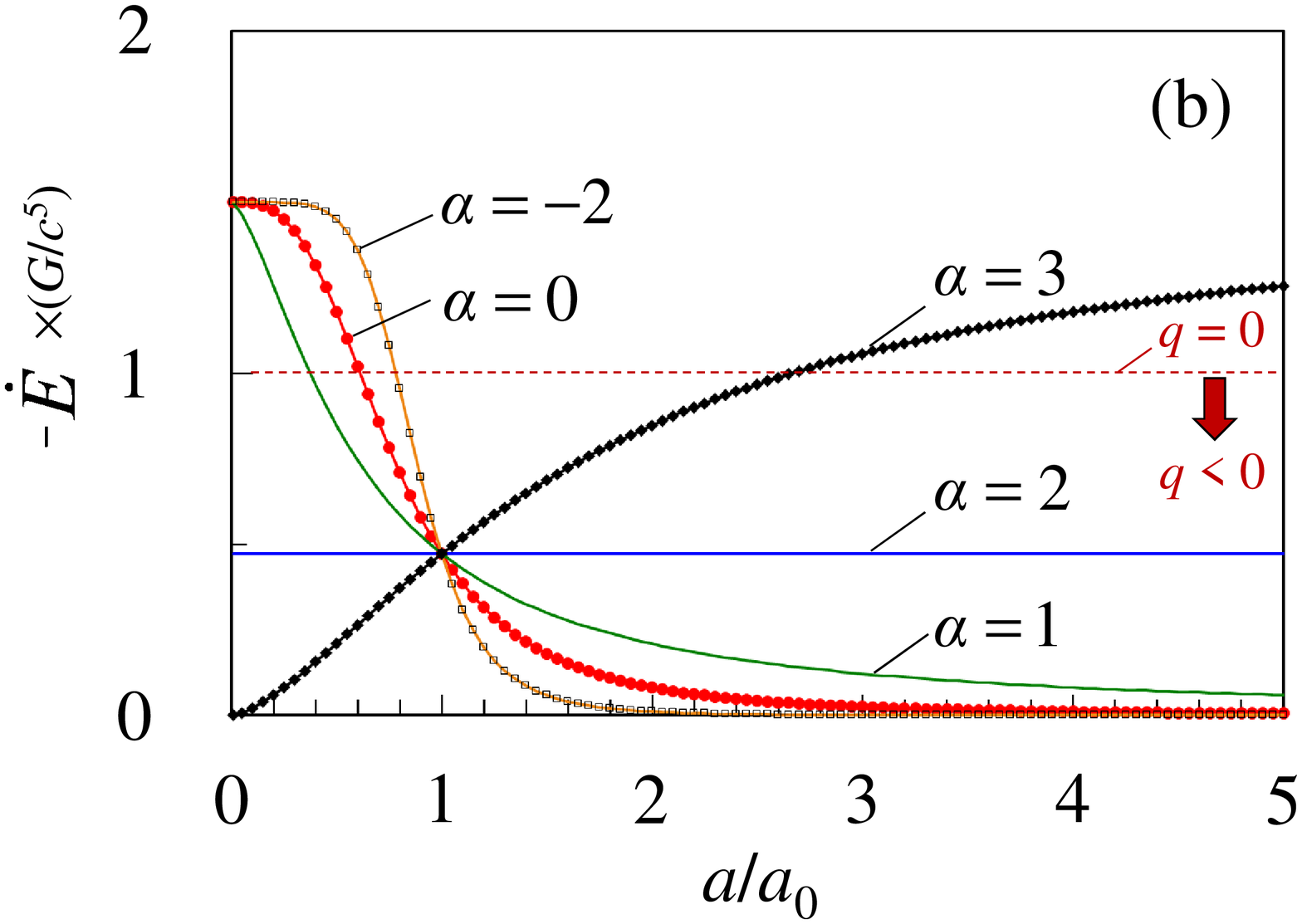}}
\end{center}
\end{minipage}
\caption{ (Color online). Evolution of the Hubble parameter and the energy flow across the horizon for the present model for $\Psi_{\alpha} =0.685$.
(a) Normalized Hubble parameter $H/H_{0}$. 
(b) Normalized energy flow $- \dot{E}  (G/c^{5})$.
$\Psi_{\alpha}$ is set to $0.685$, which is equivalent to $\Omega_{\Lambda}$ for the fine-tuned $\Lambda$CDM model. 
The plots for $\alpha =0$ are equivalent to those for the $\Lambda$CDM model.
In (b), the horizontal break line represents $- \dot{E}  (G/c^{5}) =1$, i.e., $q=0$.
The region below the break line corresponds to $q<0$, i.e., an accelerating universe, as designated by the arrow. }
\label{Fig-S_HH0_-dEdt-a_power}
\end{figure}

We now show the background evolution of the universe in the present model.
To this end, the evolution of the Hubble parameter and the energy flow is shown in Fig.\ \ref{Fig-S_HH0_-dEdt-a_power}. 
For typical results, $\alpha$ is set to $-2$, $0$, $1$, $2$, and $3$.
In addition, $\Psi_{\alpha}$ is set to $0.685$, which is equivalent to $\Omega_{\Lambda}$ for the fine-tuned $\Lambda$CDM model.
Therefore, the plots for $\alpha =0$ are equivalent to those for the $\Lambda$CDM model examined in Figs.\ \ref{Fig-H-a_1} and \ref{Fig-S_dSdt_dEdt-a_LCDM}.
As shown in Fig.\ \ref{Fig-S_HH0_-dEdt-a_power}(a), the Hubble parameter decreases with $a/a_{0}$.
For $\alpha <2$, $H/H_{0}$ gradually approaches a positive constant value, whereas for $\alpha = 3$, it gradually approaches zero.
Consequently, $\dot{H} \leq 0$ is always satisfied and $\dot{H} \rightarrow 0$ should be satisfied at least in the last stage.
In this way, $H/H_{0}$ decreases with $a/a_{0}$.
However, the evolution of the energy flow is different because $- \dot{E}$ is proportional to $- \dot{H} /H^{2}$.

As shown in Fig.\ \ref{Fig-S_HH0_-dEdt-a_power}(b), $- \dot{E}$ for $\alpha <2$ decreases with $a/a_{0}$, whereas $- \dot{E}$ for $\alpha =3$ increases with $a/a_{0}$ .
In particular, when $\alpha =2$, the energy flow is always constant (i.e., steady).
From Eq.\ (\ref{eq:-dEdt3_dEdt_ST_0_power}), the normalized constant value for $\alpha =2$ is given by
\begin{equation}
-\dot{E}  \left ( \frac{G}{ c^{5} } \right )     =   - \frac{\dot{H}}{H^{2}}  =   \frac{3 (1-\Psi_{\alpha})}{2}   .
\label{eq:-dHdt_H2_alpha2}
\end{equation}
In Fig.\ \ref{Fig-S_HH0_-dEdt-a_power}(b), the value is $0.4725$ because $\Psi_{\alpha} =0.685$.
In addition, $- \dot{E}$ for $\alpha <2$ gradually approaches zero.
This result implies $\dot{H} \rightarrow 0$ at least in the last stage.
This is because $- \dot{E}$ is proportional to $- \dot{H} /H^{2}$ and $H/H_{0}$ for $\alpha <2$ gradually approaches a positive constant value [Fig.\ \ref{Fig-S_HH0_-dEdt-a_power}(a)].
Accordingly, the last stage for $\alpha <2$ should be a kind of de Sitter universe discussed in Refs.\ \cite{Pavon2013,Mimoso2013,Krishna2017,Krishna2019}.

The energy flow is closely related to the temporal deceleration parameter $q$.
We can evaluate $q$ for the present model using Fig.\ \ref{Fig-S_HH0_-dEdt-a_power}(b) and the simple relation $q = - \dot{E}  (G/c^{5}) -1$ given by Eq.\ (\ref{eq:q_dEdt}). 
To this end, the horizontal break line of $q=0$ is plotted in Fig.\ \ref{Fig-S_HH0_-dEdt-a_power}(b).
The region below the break line corresponds to an accelerating universe because $q < 0$.

As shown in Fig.\ \ref{Fig-S_HH0_-dEdt-a_power}(b), all the energy flows are below the break line at $a/a_{0}=1$.
That is, the plots shown in this figure correspond to an accelerating universe at the present time.
In addition, an initially decelerating and then accelerating universe (hereafter `decelerating and accelerating universe') can be confirmed from the evolution of $- \dot{E}$ for $\alpha = -2$, $0$, and $1$.
This is because $- \dot{E}$ is initially larger than $1$ and thereafter smaller than $1$.
In contrast, $- \dot{E}$ for $\alpha = 3$ indicates an initially accelerating and then decelerating universe (hereafter `accelerating and decelerating universe').

In the above discussion, several cases were mentioned.
The deceleration parameter depends on $\alpha$, $\Psi_{\alpha}$, and $H/H_{0}$, as shown in Eq.\ (\ref{eq:q_power}).
Therefore, we discuss an accelerating universe using the $(\alpha, \Psi_{\alpha})$ plane.
To this end, the boundary required for $q = 0$ is calculated from Eq.\ (\ref{eq:q_power}).
Substituting $q = 0$ into Eq.\ (\ref{eq:q_power}) yields the boundary given by
\begin{align}  
   \Psi_{\alpha} = \frac{1}{3}  \left (  \frac{  H }{  H_{0} } \right )^{2 - \alpha}    . 
\label{eq:q_power is 0}
\end{align}
When $\alpha = 2$, $\Psi_{\alpha} = \frac{1}{3}$ is obtained from this equation.
When $\alpha \neq 2$, substituting Eq.\ (\ref{eq:Sol_HH0_power}) into Eq.\ (\ref{eq:q_power is 0}) yields
\begin{align}  
   \Psi_{\alpha} = \frac{1}{3} \left [ (1- \Psi_{\alpha})   \left ( \frac{a}{a_{0}} \right )^{ - \frac{3(2-\alpha)}{2}  }  + \Psi_{\alpha} \right ] , 
\label{eq:q_power is 0_aa0}
\end{align}
and solving Eq.\ (\ref{eq:q_power is 0_aa0}) with respect to $\Psi_{\alpha}$ yields the following boundary for $q = 0$:
\begin{align}  
   \Psi_{\alpha} =  \frac{  \left ( \frac{a}{a_{0}} \right )^{ - \frac{3(2-\alpha)}{2}  }       }{  2  +  \left ( \frac{a}{a_{0}} \right )^{ - \frac{3(2-\alpha)}{2}  }       }  .
\label{eq:q_power is 0_aa0_2}
\end{align}

\begin{figure} [t] 
\begin{minipage}{0.495\textwidth}
\begin{center}
\scalebox{0.34}{\includegraphics{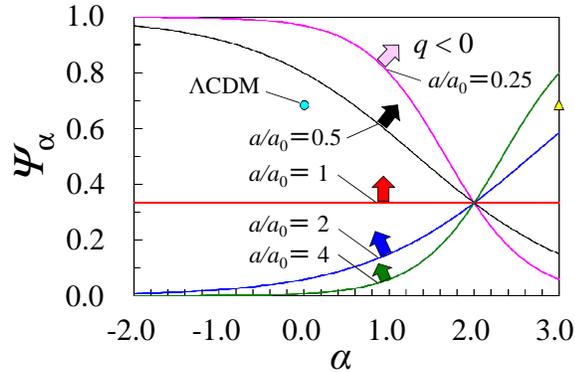}}
\end{center}
\end{minipage}
\caption{ (Color online).  Boundary of $q= 0$ in the $(\alpha, \Psi_{\alpha})$ plane for various values of $a/a_{0}$.
The boundaries for $a/a_{0}=0.25$, $0.5$, $1$, $2$, and $4$ are shown.
The arrow attached to each boundary indicates an accelerating-universe-side region that satisfies $q< 0$.
The open circle represents $(\alpha, \Psi_{\alpha}) = (0, 0.685)$ for the fine-tuned $\Lambda$CDM model.
The open triangle represents $(\alpha, \Psi_{\alpha}) = (3, 0.685)$, corresponding to the plot for $\alpha =3$ shown in Fig.\ \ref{Fig-S_HH0_-dEdt-a_power}.
The intersection point is $(\alpha, \Psi_{\alpha}) = (2, \frac{1}{3})$. }
\label{Fig-q_plane_power}
\end{figure}

Using Eq.\ (\ref{eq:q_power is 0_aa0_2}), the boundary of $q = 0$ for various values of $a/a_{0}$ can be plotted in the $(\alpha, \Psi_{\alpha})$ plane.
In Fig.\ \ref{Fig-q_plane_power}, $a/a_{0}$ is set to $0.25$, $0.5$, $1$, $2$, and $4$ to examine typical boundaries.
In an expanding universe, $a/a_{0}$ increases with time.
The arrow attached to each boundary indicates an accelerating-universe-side region that satisfies $q< 0$.
The upper side of each boundary corresponds to this region.

As shown in Fig.\ \ref{Fig-q_plane_power}, the accelerating-universe-side region varies with $a/a_{0}$.
For example, the boundary for $a/a_{0}=0.25$ in an earlier stage implies that a large-$\alpha$ and large-$\Psi_{\alpha}$ region tends to be on the accelerating universe side.
In contrast, the boundary for $a/a_{0}=4$ in a later stage implies that a small-$\alpha$ and large-$\Psi_{\alpha}$ region tends to be on the accelerating-universe side.

When $\alpha <2$, the accelerating-universe-side region extends downward with increasing $a/a_{0}$.
Consequently, a decelerating and accelerating universe is expected when $\alpha <2$.
To examine this, we focus on the point $(0, 0.685)$ for the fine-tuned $\Lambda$CDM model.
For $a/a_{0} =0.25$ and $0.5$, the point $(0, 0.685)$ is outside an accelerating-universe-side region.
However, the point is inside the region for $a/a_{0} = 1$, $2$, and $4$.
This result confirms the decelerating and accelerating universe.

In contrast, when $\alpha >2$, an accelerating and decelerating universe is expected.
To examine this, we focus on the point $(\alpha, \Psi_{\alpha}) = (3, 0.685)$, corresponding to the plot for $\alpha =3$ shown in Fig.\ \ref{Fig-S_HH0_-dEdt-a_power}.
As shown in Fig.\ \ref{Fig-q_plane_power}, the point $(3, 0.685)$ is inside the accelerating-universe-side region for $a/a_{0} =0.25$, $0.5$, $1$, and $2$.
However, for $a/a_{0} = 4$, the point is outside the region, as expected.
In this way, the dynamical properties of the present model can be systematically examined using the $(\alpha, \Psi_{\alpha})$ plane.
In the next section, we study the thermodynamic properties of the present model.

When $\alpha= 2$, energy flows are steady, as shown in Fig.\ \ref{Fig-S_HH0_-dEdt-a_power}(b) and Eq.\ (\ref{eq:-dHdt_H2_alpha2}), because a constant $\Psi_{\alpha}$ is considered.
In particular, the deceleration parameter $q$ is always zero at the intersection point $(\alpha, \Psi_{\alpha}) = (2, \frac{1}{3})$, as shown in Fig.\ \ref{Fig-q_plane_power}.
$q=0$ can be obtained by substituting $(\alpha, \Psi_{\alpha}) = (2, \frac{1}{3})$ into Eq.\ (\ref{eq:q_power}).
Accordingly, both the energy flow and the expansion of the universe are steady at the intersection point.
In this sense, the universe mentioned here is considered to be in a kind of nonequilibrium steady state.
This universe may provide new insights into discussions of nonequilibrium steady states, such as those in fluctuation theorems \cite{Evans1993etc}, thermodynamic uncertainty relations \cite{Barato2015etc}, and other related topics \cite{Sasa_Shiraishi}.

It is worth noting three assumptions used for the present model again.
(The three assumptions are closely discussed in Sec.\ \ref{Energy flow Equipartition}.)
Firstly, the equipartition law of energy can be applied to a relaxation-like process.
Secondly, the energy on the horizon can be given by Eq.\ (\ref{E_equip}).
Thirdly, the temperature given by Eq.\ (\ref{eq:T_1}) can be applied to the relaxation-like process.
The three assumptions have not been established although they were used in previous works.
In this paper, these assumptions are considered to be a viable scenario to discuss the thermodynamics on the horizon of the universe in the present model.
Note that the formulation of the present model is equivalent to that of $\Lambda(t)$CDM models although the theoretical backgrounds are different.

\section{Entropy evolution for the present model} 
\label{SL}

In this section, we study the thermodynamics on the horizon of the universe in the present model.
To this end, we examine the evolution of the Bekenstein--Hawking entropy because it is approximately equivalent to the total entropy of the late universe \cite{Egan1}.
The second law of thermodynamics is examined in Sec.\ \ref{dSdt}.
The maximization of entropy is investigated in Sec.\ \ref{dS2dt2}.

Before proceeding further, we write the present model again.
From Eq.\ (\ref{eq:Back_power_11}), the differential equation is  
\begin{align}  
    \dot{H}  & = - \frac{3}{2} H^{2}  +  \frac{3}{2}  \Psi_{\alpha} H_{0}^{2}  \left (  \frac{  H }{  H_{0} } \right )^{\alpha}        \notag \\
                 & = - \frac{3}{2} H^{2}  \left (  1 -   \Psi_{\alpha} \left (  \frac{H}{H_{0}} \right )^{\alpha -2} \right )      .  
\label{eq:Back_power_12}
\end{align}
This equation is satisfied for all $\alpha$.
The solutions can be categorized according to whether $\alpha =2$ or not.
The solution for $\alpha \neq 2$ given by Eq.\ (\ref{eq:Sol_HH0_power}) is written as 
\begin{equation}  
    \left ( \frac{H}{H_{0}} \right )^{2-\alpha}  =   (1- \Psi_{\alpha})   \left ( \frac{a}{a_{0}} \right )^{ - \frac{3(2-\alpha)}{2}  }  + \Psi_{\alpha}       , 
\label{eq:Sol_HH0_power_22}
\end{equation}
and the solution for $\alpha = 2$ given by Eq.\ (\ref{eq:Sol_HH0_aa0_H2}) is  
\begin{equation}
     \frac{H}{H_{0}}  =     \left ( \frac{a}{a_{0}} \right )^{ - \frac{3 (1- \Psi_{\alpha}) }{2}  }     .
\label{eq:Sol_HH0_aa0_H2_2}
\end{equation}

From the above solutions, $S_{\rm{BH}}$ for the present model is calculated.
When $\alpha \neq 2$, substituting Eq.\ (\ref{eq:Sol_HH0_power_22}) into $S_{\rm{BH}} = K/H^{2}$ given by Eq.\ (\ref{eq:SBH2}) yields
\begin{align}  
S_{\rm{BH}}   
&=  \frac{K}{H^{2}}  =  \frac{K / H_{0}^{2} }{ (H/H_{0})^{2}   }                        \notag \\
&=  \frac{K / H_{0}^{2} }{ \left (     (1- \Psi_{\alpha})   \left ( \frac{a}{a_{0}} \right )^{ - \frac{3(2-\alpha)}{2}  }  + \Psi_{\alpha}     \right )^{\frac{2}{2-\alpha}}     }     \notag \\
&=  \frac{K}{H_{0}^{2}}     \left (     (1- \Psi_{\alpha})   \left ( \frac{a}{a_{0}} \right )^{ - \frac{3(2-\alpha)}{2}  }  + \Psi_{\alpha}     \right )^{\frac{2}{\alpha-2}}      ,
\label{eq:SBH_power_Complicate}      
\end{align}  
where $K$ is a positive constant given by Eq.\ (\ref{eq:K-def}).
When $\alpha =2$, applying Eq.\ (\ref{eq:Sol_HH0_aa0_H2_2}) instead of Eq.\ (\ref{eq:Sol_HH0_power_22}) yields
\begin{align}  
S_{\rm{BH}}   &=   \frac{K / H_{0}^{2}  }{ \left ( \frac{a}{a_{0}} \right )^{ - 3 (1- \Psi_{\alpha})}   }  = \frac{K}{H_{0}^{2}}  \left ( \frac{a}{a_{0}} \right )^{ 3 (1- \Psi_{\alpha})  }   .  
\label{eq:SBH_power_Complicate_H2}      
\end{align}  
The obtained $S_{\rm{BH}}$ is positive.

\subsection{Second law of thermodynamics: $\dot{S}_{\rm{BH}} \geq 0$}
\label{dSdt}

In this subsection, we examine the second law of thermodynamics on the Hubble horizon for the present model.
From Eqs.\ (\ref{eq:dSBH}) and (\ref{eq:dSBH_2}), the first derivative of $S_{\rm{BH}}$ is written as
\begin{equation}
\dot{S}_{\rm{BH}}   =  \frac{-2K \dot{H} }{H^{3}}  =  2 S_{\rm{BH}} \left ( \frac{- \dot{H} }{H} \right )                .
\label{eq:dSBH_2_3}      
\end{equation}
This equation indicates that the sign of $\dot{S}_{\rm{BH}} $ depends on whether $- \dot{H} /H $ is positive or negative because $ S_{\rm{BH}} > 0$.

We now calculate $\dot{S}_{\rm{BH}}$ for the present model.
When $\alpha \neq 2$, substituting Eq.\ (\ref{eq:Back_power_12}) into Eq.\ (\ref{eq:dSBH_2_3}) and applying Eq.\ (\ref{eq:Sol_HH0_power_22}) yields
\begin{align}  
\dot{S}_{\rm{BH}}   
&=  \frac{-2K \dot{H} }{H^{3}}  =  \frac{2K}{H_{0}}  \left ( \frac{- \dot{H} }{H^{2}} \right )    \frac{H_{0}}{H}                                    \notag \\
&=  \frac{2K}{H_{0}}  \frac{3}{2} \left (  1 -   \Psi_{\alpha} \left (  \frac{H}{H_{0}} \right )^{\alpha -2}  \right )      \frac{H_{0}}{H}     \notag \\
&=  \frac{3K}{H_{0}}  \left (  1 -   \frac{ \Psi_{\alpha} }{ (1- \Psi_{\alpha})   \left ( \frac{a}{a_{0}} \right )^{ - \frac{3(2-\alpha)}{2}  }  + \Psi_{\alpha}    }  \right )       \notag \\
& \quad \times  \left [ (1- \Psi_{\alpha})   \left ( \frac{a}{a_{0}} \right )^{ - \frac{3(2-\alpha)}{2}  }  + \Psi_{\alpha}  \right ]^{ \frac{1}{\alpha-2} }  .
\label{eq:dSBH_2_3_power_Complicate}      
\end{align}  
When $\alpha =2$, applying Eq.\ (\ref{eq:Sol_HH0_aa0_H2_2}) instead of Eq.\ (\ref{eq:Sol_HH0_power_22}) yields
\begin{align}  
\dot{S}_{\rm{BH}}   
&=  \frac{2K}{H_{0}}  \left ( \frac{- \dot{H} }{H^{2}} \right )    \frac{H_{0}}{H}          =  \frac{2K}{H_{0}}  \frac{3}{2}  (  1 -   \Psi_{\alpha}  )      \left ( \frac{a}{a_{0}} \right )^{ \frac{3 (1- \Psi_{\alpha}) }{2}  }          \notag \\
&=  \frac{3K}{H_{0}}  (  1 -   \Psi_{\alpha}  )      \left ( \frac{a}{a_{0}} \right )^{ \frac{3 (1- \Psi_{\alpha}) }{2}  }       .
\label{eq:dSBH_2_3_power_Complicate_H2}      
\end{align}  
Although the obtained equations are slightly complicated, they can be used to calculate
the evolution of $\dot{S}_{\rm{BH}}$.

In the present model, we assume $ 0 \leq \Psi_{\alpha} \leq 1 $, as shown in Eq.\ (\ref{eq:psi0}).
Therefore, Eq.\ (\ref{eq:dSBH_2_3_power_Complicate_H2}) for $\alpha =2$ always satisfies $\dot{S}_{\rm{BH}} \geq 0$.
In addition, applying $ 0 \leq \Psi_{\alpha} \leq 1 $ to Eq.\ (\ref{eq:Sol_HH0_power_22}) for $\alpha \neq 2$ yields
\begin{equation}  
    \left ( \frac{H}{H_{0}} \right )^{2-\alpha}  =      (1- \Psi_{\alpha})   \left ( \frac{a}{a_{0}} \right )^{ - \frac{3(2-\alpha)}{2}  }  + \Psi_{\alpha}         \geq   \Psi_{\alpha}     \geq     0        ,
\label{eq:Sol_HH0_power_22_2}
\end{equation}
and rearranging the above inequality gives
\begin{equation}  
    1 -   \frac{ \Psi_{\alpha} }{ (1- \Psi_{\alpha})   \left ( \frac{a}{a_{0}} \right )^{ - \frac{3(2-\alpha)}{2}  }  + \Psi_{\alpha}    }   \geq   0    .
\label{eq:Sol_power_100}
\end{equation}
Using the two inequalities, we can confirm that Eq.\ (\ref{eq:dSBH_2_3_power_Complicate}) for $\alpha \neq 2$ always satisfies $\dot{S}_{\rm{BH}} \geq 0$.
Therefore, the second law of thermodynamics on the horizon is satisfied in the present model.
This result leads to $ \dot{H} /H  \leq 0$ because $\dot{S}_{\rm{BH}} = 2 S_{\rm{BH}} (- \dot{H} /H)$ given by Eq.\ (\ref{eq:dSBH_2_3}).
 
The present model satisfies $\dot{S}_{\rm{BH}} \geq 0$ and $ \dot{H} /H \leq 0$ without assuming the sign of $H$.
However, observations indicate $H >0$ \cite{Hubble2017}, as shown in Fig.\ \ref{Fig-H-a_1}.
Accordingly, in this paper, we assume an expanding universe, i.e., $H > 0$.
Consequently, $\dot{H} \leq 0$ is obtained from $ \dot{H} /H  \leq 0$.
$\dot{H} \leq 0$ is consistent with the observed data points shown in Fig.\ \ref{Fig-H-a_1}.

In summary, the present model always satisfies the second law of thermodynamics on the horizon:
\begin{equation} 
 \dot{S}_{\rm{BH}}   \geq 0    \quad \textrm{and} \quad   \frac{ \dot{H} }{H}  \leq 0        ,
\label{eq:dSdt_power_0}
\end{equation}
where $ 0 \leq \Psi_{\alpha} \leq 1 $ is assumed.
In addition, assuming an expanding universe, i.e., $H >0$, we have
\begin{equation} 
 \dot{H}               \leq 0  .
\label{eq:-dotH_power_0}
\end{equation}

\subsection{Maximization of entropy: $\ddot{S}_{\rm{BH}} < 0$}
\label{dS2dt2}

We discuss the maximization of the entropy on the Hubble horizon of an expanding universe in the present model.
From Eq.\ (\ref{eq:d2SB_1}), the second derivative of $S_{\rm{BH}}$ is written as
\begin{equation}  
\ddot{S}_{\rm{BH}}   =  2 S_{\rm{BH}}       \left ( \frac{   3 \dot{H}^{2}   - \ddot{H} H  }{H^{2}}  \right )     .
\label{eq:d2SB_1_11}      
\end{equation}
Equation\ (\ref{eq:d2SB_1_11}) indicates that the sign of $\ddot{S}_{\rm{BH}}$ depends on whether the $3 \dot{H}^{2}   - \ddot{H} H$ term is positive or negative.

To examine the sign of this term, we calculate this term from Eq.\ (\ref{eq:Back_power_12}).
The detail calculation is given in Appendix \ref{Calculation2}.
From Eq.\ (\ref{eq:B1set5_A}), $3 \dot{H}^{2}   - \ddot{H} H $ is written as
\begin{align}  
   3 \dot{H}^{2}   - \ddot{H} H     &=  \frac{3}{2} ( - \dot{H} ) H^{2}    \left [ 1 -  \Psi_{\alpha}  (3 - \alpha )    \left (  \frac{  H }{  H_{0} } \right )^{\alpha-2}   \right ]   .
\label{eq:B1set5_20}
\end{align}
This equation is satisfied for all $\alpha$.
Substituting Eq.\ (\ref{eq:B1set5_20}) into Eq.\ (\ref{eq:d2SB_1_11}) yields
\begin{align}  
\ddot{S}_{\rm{BH}}   
                              &=2 S_{\rm{BH}}  \left ( \frac{  \frac{3}{2}     ( - \dot{H} ) H^{2}   \left [ 1 -  \Psi_{\alpha}  (3 - \alpha )    \left (  \frac{  H }{  H_{0} } \right )^{\alpha-2}   \right ]     }{H^{2}}  \right )   \notag \\
                              &=3 S_{\rm{BH}}    ( - \dot{H} )   \left [ 1 -  \Psi_{\alpha}  (3 - \alpha )    \left (  \frac{  H }{  H_{0} } \right )^{\alpha-2}   \right ]   ,
\label{eq:d2Sdt2_power_1}
\end{align}
and applying $S_{\rm{BH}} = K/H^{2}$ given by Eq.\ (\ref{eq:SBH2}) yields
\begin{align}  
\ddot{S}_{\rm{BH}}   &=3 K \left (  \frac{ - \dot{H} }{H^{2}}  \right )          \left [ 1 -  \Psi_{\alpha}  (3 - \alpha )    \left (  \frac{  H }{  H_{0} } \right )^{\alpha-2}   \right ]   .
\label{eq:d2Sdt2_power_1B}
\end{align}
When $\alpha =2$, a simple relation can be obtained from Eq.\ (\ref{eq:d2Sdt2_power_1B}).
Substituting $ -\dot{H} / H^{2}$ for $\alpha =2$ given by Eq.\ (\ref{eq:-dHdt_H2_alpha2}) into Eq.\ (\ref{eq:d2Sdt2_power_1B}) and applying $\alpha =2$ yields
\begin{align}  
\ddot{S}_{\rm{BH}}  &=  3 K  \frac{3 (1-\Psi_{\alpha})}{2}  (1-  \Psi_{\alpha}  )      
                               =   \frac{9 K}{2} (1-\Psi_{\alpha})^{2}    . 
\label{eq:d2Sdt2_power_1B_alpha2}
\end{align}
Equation\ (\ref{eq:d2Sdt2_power_1B_alpha2}) indicates that $\ddot{S}_{\rm{BH}} < 0$ is not satisfied when $\alpha = 2$. 
When $\alpha \neq 2$, substituting Eq.\ (\ref{eq:Back_power_12}) into Eq.\ (\ref{eq:d2Sdt2_power_1B}) and applying Eq.\ (\ref{eq:Sol_HH0_power_22}) to the resultant equation yields the following slightly complicated equation:
\begin{align}  
& \ddot{S}_{\rm{BH}}   
  =3 K \left (  \frac{ - \dot{H} }{H^{2}}  \right )          \left [ 1 -  \Psi_{\alpha}  (3 - \alpha )    \left (  \frac{  H }{  H_{0} } \right )^{\alpha-2}   \right ]                \notag \\
&= \frac{9K}{2}  \left (  1 -   \Psi_{\alpha} \left (  \frac{H}{H_{0}} \right )^{\alpha -2} \right )   \left [ 1 -  \Psi_{\alpha}  (3 - \alpha )    \left (  \frac{  H }{  H_{0} } \right )^{\alpha-2}   \right ]   \notag \\
&= \frac{9K}{2}   \left (  1 -  \frac{  \Psi_{\alpha}                      }{   (1- \Psi_{\alpha})   \left ( \frac{a}{a_{0}} \right )^{ - \frac{3(2-\alpha)}{2}  }  + \Psi_{\alpha}         }  \right )    \notag \\
& \quad \times           \left [  1 -   \frac{  \Psi_{\alpha}  (3 - \alpha )  }{   (1- \Psi_{\alpha})   \left ( \frac{a}{a_{0}} \right )^{ - \frac{3(2-\alpha)}{2}  }  + \Psi_{\alpha}         }  \right ]   .
\label{eq:d2Sdt2_power_1_Complicate}
\end{align}
This equation reduces to Eq.\ (\ref{eq:d2Sdt2_power_1B_alpha2}) when $\alpha \rightarrow 2$.

\begin{figure}[p] 
\begin{minipage}{0.495\textwidth}
\begin{center}
\scalebox{0.34}{\includegraphics{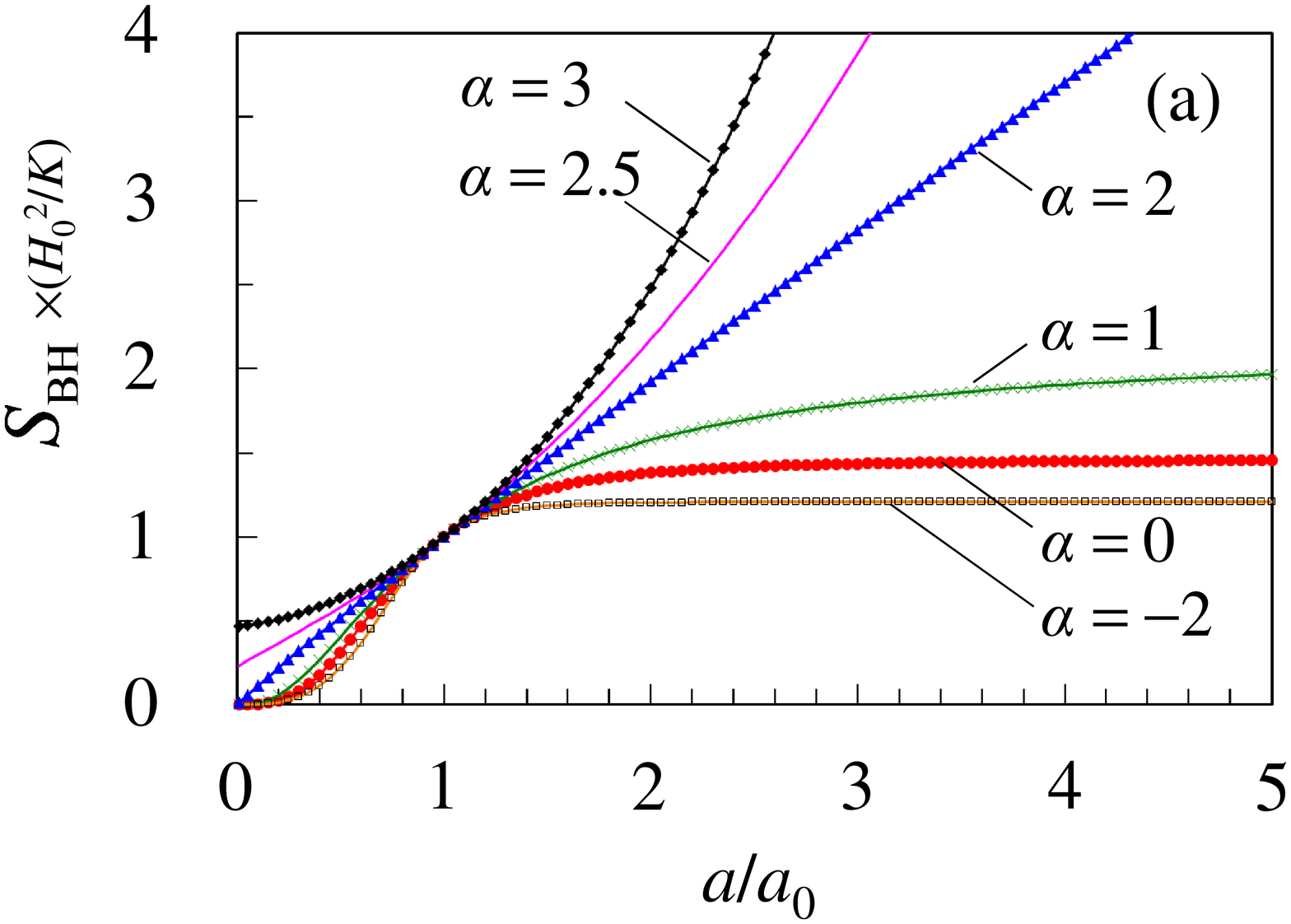}}
\end{center}
\end{minipage}
\begin{minipage}{0.495\textwidth}
\begin{center}
\scalebox{0.34}{\includegraphics{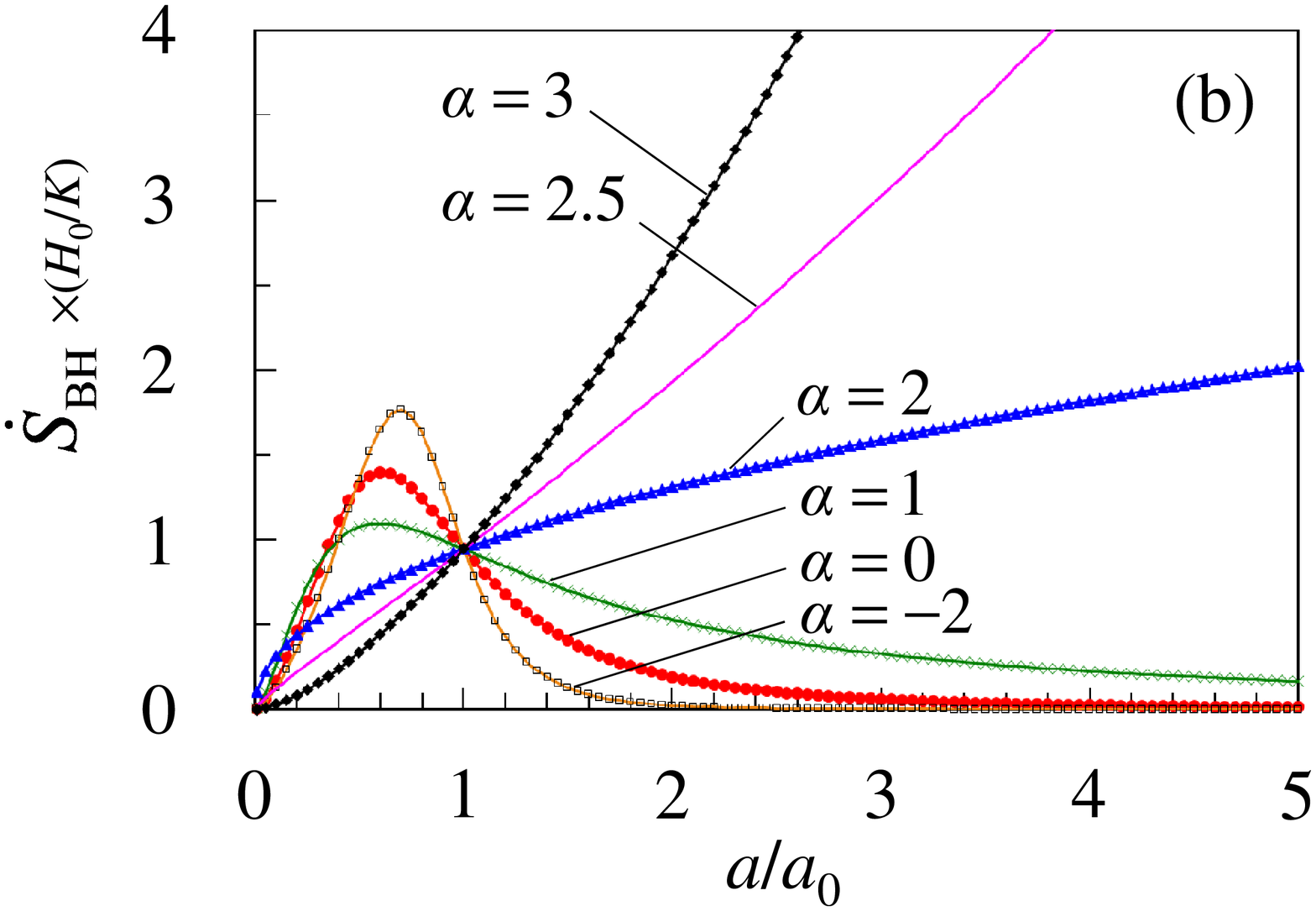}}
\end{center}
\end{minipage}
\begin{minipage}{0.495\textwidth}
\begin{center}
\scalebox{0.34}{\includegraphics{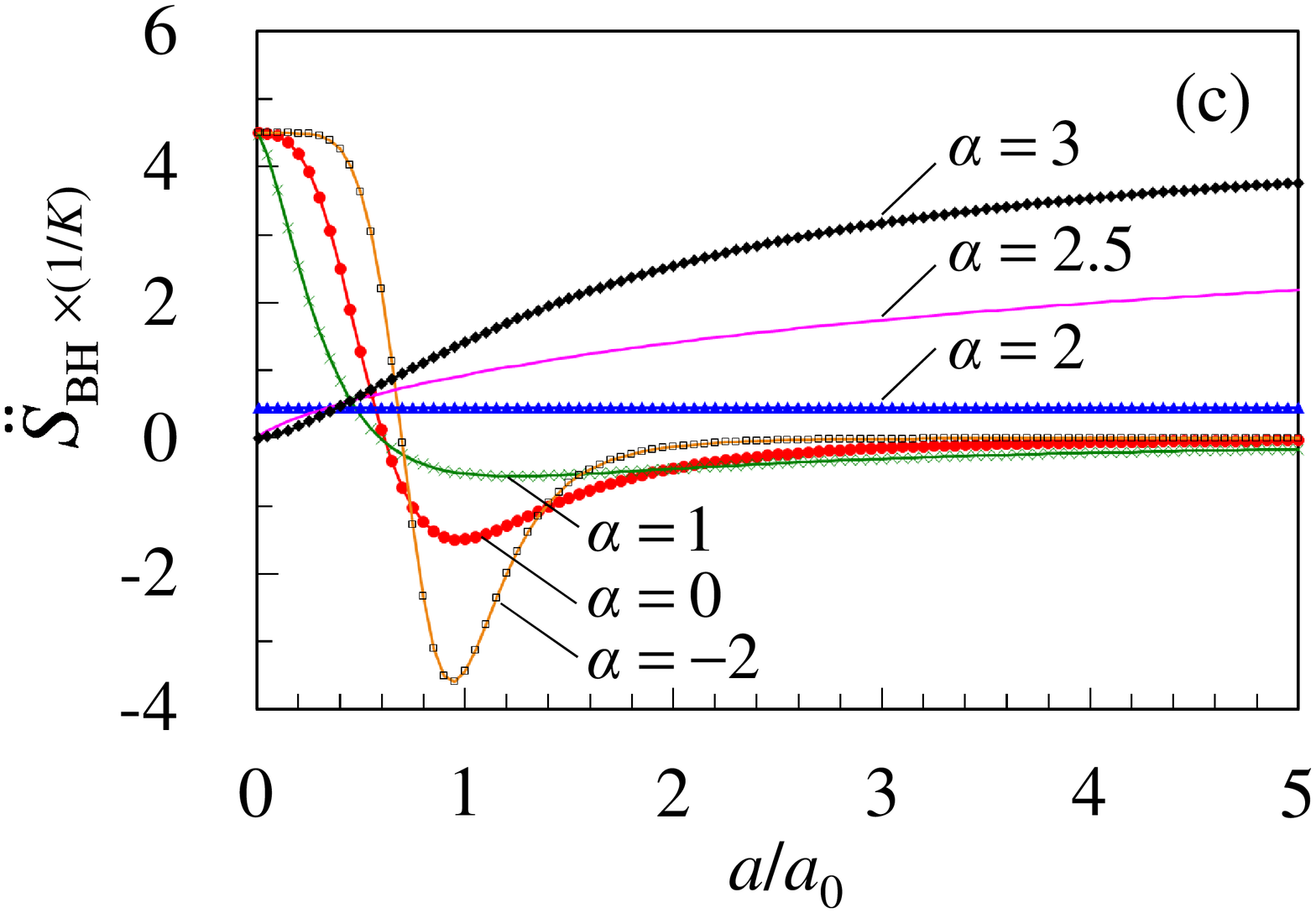}}
\end{center}
\end{minipage}
\caption{ (Color online). Evolution of three normalized entropic parameters for the present model for $\Psi_{\alpha} =0.685$.
(a) $S_{\rm{BH}}$ normalized as $S_{\rm{BH}} (H_{0}^{2}/K)$. (b) $\dot{S}_{\rm{BH}}$ normalized as $\dot{S}_{\rm{BH}} (H_{0}/K)$. (c) $\ddot{S}_{\rm{BH}}$ normalized as $\ddot{S}_{\rm{BH}} (1/K)$. 
The plots for $\alpha =0$ are equivalent to those for the fine-tuned $\Lambda$CDM model because $\Psi_{\alpha} =0.685$.}
\label{Fig-S_dSdt_d2Sdt2-a_power}
\end{figure}

We now show the evolution of three entropic parameters, namely $S_{\rm{BH}}$, $\dot{S}_{\rm{BH}}$, and $\ddot{S}_{\rm{BH}}$, for the present model. 
To examine typical results, $\alpha$ is set to $-2$, $0$, $1$, $2$, $2.5$, and $3$. 
In addition, $\Psi_{\alpha}$ is set to $0.685$, which is equivalent to $\Omega_{\Lambda}$ for the fine-tuned $\Lambda$CDM model.  

As shown in Fig.\ \ref{Fig-S_dSdt_d2Sdt2-a_power}(a), $S_{\rm{BH}}$ increases with $a/a_{0}$.
When $\alpha$ is positive, $S_{\rm{BH}}$ is strongly influenced by $\alpha$.
In contrast, when $\alpha$ is negative, $S_{\rm{BH}}$ is weakly  influenced by $\alpha$.
For all $\alpha$, $S_{\rm{BH}}$ increases with $a/a_{0}$.
Therefore, $\dot{S}_{\rm{BH}}$ is positive, as shown in Fig.\ \ref{Fig-S_dSdt_d2Sdt2-a_power}(b).
This result agrees with Eq.\ (\ref{eq:dSdt_power_0}). 
In addition, $\dot{S}_{\rm{BH}}$ for $\alpha <2$ increases with $a/a_{0}$ in the early stage ($a/a_{0} \lessapprox 0.6$).
However, it gradually decreases with $a/a_{0}$ in the last stage ($1 \ll a/a_{0}$).
Consequently, $\ddot{S}_{\rm{BH}}$ for $\alpha < 2$ is positive in the early stage and negative in the last stage [Fig.\ \ref{Fig-S_dSdt_d2Sdt2-a_power}(c)].
In contrast, $\dot{S}_{\rm{BH}}$ for $\alpha \geq 2$ increases with $a/a_{0}$ [Fig.\ \ref{Fig-S_dSdt_d2Sdt2-a_power}(b)].
Accordingly, $\ddot{S}_{\rm{BH}}$ for $\alpha \geq 2$ is likely positive even in the last stage [Fig.\ \ref{Fig-S_dSdt_d2Sdt2-a_power}(c)].
When $\alpha =2$, $\ddot{S}_{\rm{BH}}$ is constant.
This is because $\ddot{S}_{\rm{BH}}$ for $\alpha =2$ is given by Eq.\ (\ref{eq:d2Sdt2_power_1B_alpha2}) and $\Psi_{\alpha}$ is constant.
In Fig.\ \ref{Fig-S_dSdt_d2Sdt2-a_power}(c), the normalized value of $\ddot{S}_{\rm{BH}}$ is approximately $0.447$, as calculated from Eq.\ (\ref{eq:d2Sdt2_power_1B_alpha2}) and $\Psi_{\alpha} =0.685$.

These results imply that the maximization of entropy is not satisfied when $\alpha \ge 2$, but should be satisfied in the last stage when $\alpha <2$.
So far, typical results have been observed. 
Next, we systematically examine entropy maximization. 
In the present model, $\dot{H} \leq 0$ is satisfied in an expanding universe, as shown in Eq.\ (\ref{eq:-dotH_power_0}).
For simplicity, we consider $ \dot{H} < 0$.
Consequently, from Eq.\ (\ref{eq:d2Sdt2_power_1B}), to satisfy $\ddot{S}_{\rm{BH}} < 0$, we require
\begin{align}  
   1 -  \Psi_{\alpha}  (3 - \alpha )    \left (  \frac{  H }{  H_{0} } \right )^{\alpha-2}   < 0  .
\label{eq:d2Sdt2 less than 0}
\end{align}
The left-hand side of Eq.\ (\ref{eq:d2Sdt2 less than 0}) is set to $F$, which is written as
\begin{align}  
  F = 1 -  \Psi_{\alpha}  (3 - \alpha )    \left (  \frac{  H }{  H_{0} } \right )^{\alpha-2}   .
\label{eq:F0}
\end{align}
Therefore, $F < 0$ is required to satisfy $\ddot{S}_{\rm{BH}} < 0$.

We now focus on the last phase in the last stage, corresponding to $ a/a_{0} \rightarrow \infty$.
To this end, we use a relation obtained from the inverse of Eq.\ (\ref{eq:Sol_HH0_power_22_2}) for $\alpha \neq 2$.
When $\alpha < 2$, applying $ a/a_{0} \rightarrow \infty$, the relation can be approximately written as
\begin{equation}  
    \left ( \frac{H}{H_{0}} \right )^{\alpha-2}  = \frac{1} {  (1- \Psi_{\alpha})   \left ( \frac{a}{a_{0}} \right )^{ - \frac{3(2-\alpha)}{2}  }  + \Psi_{\alpha}     }    \approx  \frac{1}{  \Psi_{\alpha}     }  . 
\label{eq:Sol_HH0_power_22_2_Inv}
\end{equation}
Substituting the above equation into Eq.\ (\ref{eq:F0}) yields
\begin{align}  
  F  &=       1 -  \Psi_{\alpha}  (3 - \alpha )    \left (  \frac{  H }{  H_{0} } \right )^{\alpha-2}    \notag \\
     &\approx    1 -  \Psi_{\alpha}  (3 - \alpha )   \frac{1}{  \Psi_{\alpha}     } = \alpha -2 .
\label{eq:F2}
\end{align}
This equation implies that $F < 0$ is approximately satisfied when $\alpha -2 < 0$.
Accordingly, when $ a/a_{0} \rightarrow \infty$, to satisfy $ \ddot{S}_{\rm{BH}} < 0  $, we require
\begin{align}  
  \alpha < 2    ,
\label{eq:alpha_1}
\end{align}
where $H >0$ and $ 0 \leq \Psi_{\alpha} \leq 1 $ are assumed.

As discussed above, the universe for $\alpha < 2$ is expected to approach a kind of equilibrium state at least in the last stage.
The evolution of the universe is considered to be a relaxation-like process.
Therefore, a region that satisfies $ \ddot{S}_{\rm{BH}} < 0$ in the $(\alpha, \Psi_{\alpha})$ plane varies with time before the last stage.
To examine the relaxation-like process systematically, the boundary required for $\ddot{S}_{\rm{BH}} = 0$ is calculated from Eq.\ (\ref{eq:d2Sdt2_power_1B}).
The boundary is given by
\begin{align}  
   \Psi_{\alpha} = \frac{1}{3 - \alpha }    \left (  \frac{  H }{  H_{0} } \right )^{2-\alpha}   .
\label{eq:d2Sdt2 is 0}
\end{align}
When $\alpha = 2$, $\Psi_{\alpha} = 1$ is obtained from this equation.
When $\alpha \neq 2$, substituting Eq.\ (\ref{eq:Sol_HH0_power_22}) into Eq.\ (\ref{eq:d2Sdt2 is 0}) yields
\begin{align}  
   \Psi_{\alpha} = \frac{1}{3 - \alpha } \left [ (1- \Psi_{\alpha})   \left ( \frac{a}{a_{0}} \right )^{ - \frac{3(2-\alpha)}{2}  }  + \Psi_{\alpha} \right ] , 
\label{eq:d2Sdt2_power_0_aa0}
\end{align}
and solving Eq.\ (\ref{eq:d2Sdt2_power_0_aa0}) with respect to $\Psi_{\alpha}$ yields the following boundary required for $\ddot{S}_{\rm{BH}} = 0$:
\begin{align}  
   \Psi_{\alpha} =  \frac{  \left ( \frac{a}{a_{0}} \right )^{ - \frac{3(2-\alpha)}{2}  }       }{  2 - \alpha +  \left ( \frac{a}{a_{0}} \right )^{ - \frac{3(2-\alpha)}{2}  }       }  .
\label{eq:d2Sdt2_power_0_aa0_2}
\end{align}

\begin{figure} [t] 
\begin{minipage}{0.495\textwidth}
\begin{center}
\scalebox{0.34}{\includegraphics{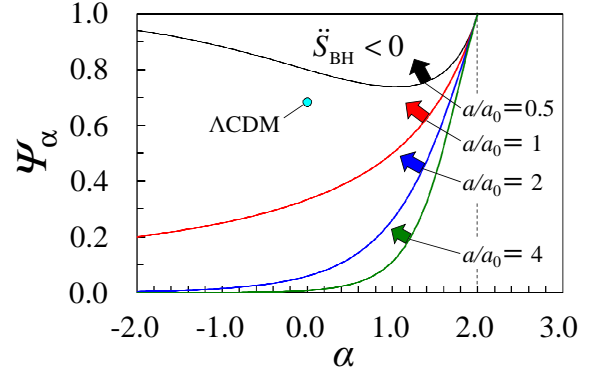}}
\end{center}
\end{minipage}
\caption{ (Color online).  Boundary of $\ddot{S}_{\rm{BH}} = 0$ in the $(\alpha, \Psi_{\alpha})$ plane for various values of $a/a_{0}$.
The four boundaries for $a/a_{0}=0.5$, $1$, $2$, and $4$ are shown.
The arrow attached to each boundary indicates a relaxation-like-process-side region that satisfies $\ddot{S}_{\rm{BH}} < 0$.
The open circle represents $(\alpha, \Psi_{\alpha}) = (0, 0.685)$ for the fine-tuned $\Lambda$CDM model. 
The vertical break line represents $\alpha =2$. }
\label{Fig-dS2dt2_plane_power}
\end{figure}

Using Eq.\ (\ref{eq:d2Sdt2_power_0_aa0_2}), the boundary of $\ddot{S}_{\rm{BH}} = 0$ for various values of $a/a_{0}$ can be plotted in the $(\alpha, \Psi_{\alpha})$ plane.
In Fig.\ \ref{Fig-dS2dt2_plane_power}, $a/a_{0}$ is set to $0.5$, $1$, $2$, and $4$ to observe typical boundaries.
The arrow attached to each boundary indicates a relaxation-like-process-side region that satisfies $\ddot{S}_{\rm{BH}} < 0$.
The upper side of each boundary corresponds to this region.
As shown in Fig.\ \ref{Fig-dS2dt2_plane_power}, this region gradually extends downward with increasing $a/a_{0}$. 
However, the region does not exceed the vertical break line of $\alpha =2$, although all the boundaries intersect at the point $(\alpha, \Psi_{\alpha}) = (2, 1)$.
This constraint is consistent with Eq.\ (\ref{eq:alpha_1}).

We now show the boundary for $a/a_{0} =1$, which represents the present time.
As shown in Fig.\ \ref{Fig-dS2dt2_plane_power}, a small-$\alpha$ and large-$\Psi_{\alpha}$ region tends to satisfy $\ddot{S}_{\rm{BH}} < 0$ at the present time.
For example, the point $(0, 0.685)$ for the fine-tuned $\Lambda$CDM model satisfies $\ddot{S}_{\rm{BH}} < 0$.
A region close to this point satisfies the maximization of entropy not only at the present time but also in the future.

As observed above, a small-$\alpha$ and large-$\Psi_{\alpha}$ region tends to approach thermodynamic equilibrium-like states relatively quickly.
In contrast, in a large-$\alpha$ and small-$\Psi_{\alpha}$ region, it should take a very long time to satisfy $\ddot{S}_{\rm{BH}} < 0$, even when $\alpha <2$.
This result implies that cosmological models in the former region should be favored from a thermodynamics viewpoint.

\section{Conclusions}
\label{Conclusions}

We examined the thermodynamics on the Hubble horizon of a flat FRW universe at late times using holographic cosmological models with a power-law term.
These models are generally categorized into two types, i.e., $\Lambda(t)$ and BV. 
In the present study, we derived energy flows across the horizon using two methods to examine their suitability.
It was found that $\Lambda(t)$ models similar to $\Lambda (t)$CDM models satisfy the consistency of the two energy flows.

Based on this consistency, we phenomenologically formulated a $\Lambda(t)$ model that includes $H^{\alpha}$ terms.
The differential equation for the present model can be analytically solved.
Using this solution, we examined the dynamic properties of the present model.
It was found that $\alpha <2$ corresponds to a decelerating and accelerating universe, whereas $\alpha >2$ corresponds to an accelerating and decelerating universe.

In addition, we examined the thermodynamic properties of the present model, focusing on the evolution of the Bekenstein--Hawking entropy $S_{\rm{BH}}$.
When $\alpha$ is positive, the entropy evolution is strongly influenced by $\alpha$.
However, when $\alpha$ is negative, $S_{\rm{BH}}$ is weakly  influenced by $\alpha$.
The present model always satisfies the second law of thermodynamics on the horizon, i.e., $\dot{S}_{\rm{BH}} \geq 0$.
When $\alpha < 2$, the maximization of the entropy, $ \ddot{S}_{\rm{BH}} < 0$, should be satisfied at least in the last stage of the evolution of an expanding universe.

Moreover, we systematically examined the relaxation-like process of the present model using the $(\alpha, \Psi_{\alpha})$ plane. 
A small-$\alpha$ and large-$\Psi_{\alpha}$ region tends to approach thermodynamic equilibrium-like states relatively quickly.
Cosmological models in this region are likely favored from a thermodynamics viewpoint.
The present study should provide new insights into various cosmological models, especially for $\Lambda (t)$CDM models.

In this paper, we focused on entropy maximization related to equilibrium-like states.
Accordingly, a universe for $(\alpha, \Psi_{\alpha}) = (2, \frac{1}{3})$, which is in a kind of nonequilibrium steady state, was not examined here.
This universe may facilitate the examination of nonequilibrium steady states in an expanding universe.
In addition, density perturbations related to structure formations were not discussed here.
Detailed studies are needed from various viewpoints.
These tasks are left for future research.

\begin{acknowledgements}
The present study was supported by JSPS KAKENHI Grant Number JP18K03613.
The author wishes to thank the anonymous referee for very valuable comments which improved the paper.
\end{acknowledgements}

\appendix

\section{Calculation of energy flow across the Hubble horizon}
\label{Solution-dEdt}

In this appendix, we calculate the right-hand side of Eq.\ (\ref{eq:-dEdt}) to derive the energy flow across the Hubble horizon of a flat FRW universe.
The energy flow given by Eq.\ (\ref{eq:-dEdt}) is written as
\begin{align}
  -\dot{E} &= A_{H} r_{H} c^{2} (1+w) \rho H    .
\label{eq:-dEdt_A}
\end{align}
In order to calculate the right-hand side of this equation, $(1+w) \rho H$ is calculated from the general continuity equation.
Using Eq.\ (\ref{eq:drho_General_00}), $(1+w) \rho H$ can be written as
\begin{equation}
        (1+w)  \rho H  =  - \frac{1}{3}\dot{\rho} +   \frac{H}{4 \pi G}  \left(  h_{\textrm{B}}(t)  -  \frac{\dot{f}_{\Lambda}(t)}{2 H }      \right )                   .
\label{eq:rhoH_A}
\end{equation}
Next, we calculate $ - \dot{\rho} /3$ in Eq.\ (\ref{eq:rhoH_A}) from the general Friedmann equation.  
Differentiating Eq.\ (\ref{eq:General_FRW01_f_0}) with respect to $t$ gives 
\begin{equation}
    2 H \dot{H}      =  \frac{ 8\pi G }{ 3 } \dot{\rho}    + \dot{f}_{\Lambda}(t)            ,                                               
\label{eq:drhodt1_A} 
\end{equation} 
and solving Eq.\ (\ref{eq:drhodt1_A}) with respect to $-\dot{\rho} /3$ yields
\begin{equation}
  - \frac{1}{3}\dot{\rho} =  \frac{- 2 H \dot{H}    + \dot{f}_{\Lambda}  } {8\pi G }          .                                                 
\label{eq:drhodt2_A} 
\end{equation} 
We now calculate the right-hand side of Eq.\ (\ref{eq:-dEdt_A}).
Substituting Eq.\ (\ref{eq:rhoH_A}) and $r_{H} = c/H$ into Eq.\ (\ref{eq:-dEdt_A}) and applying Eq.\ (\ref{eq:drhodt2_A}) yields
\begin{align}
-\dot{E}  &= A_{H} \left ( \frac{c}{H}  \right )  c^{2}  \left [  - \frac{1}{3}\dot{\rho} +   \frac{H}{4 \pi G}  \left(  h_{\textrm{B}}(t)  -  \frac{\dot{f}_{\Lambda}(t)}{2 H }      \right )   \right ]      \notag \\
             &= \frac{A_{H} c^{3}}{H}   \left [  \frac{- 2 H \dot{H}    + \dot{f}_{\Lambda} } {8\pi G }             +   \frac{H}{4 \pi G}  \left(  h_{\textrm{B}}(t)  -  \frac{\dot{f}_{\Lambda} }{2 H }      \right )    \right ]   \notag \\
             &=  \frac{ A_{H} c^{3} }{4 \pi G }  \left (  -  \dot{H}    +  h_{\textrm{B}}(t)  \right )    .
\label{eq:-dEdt2_A}
\end{align}
In addition, substituting $A_{H}=4 \pi r_{H}^2 =4 \pi (c/H)^{2} $ into Eq.\ (\ref{eq:-dEdt2_A}) yields 
\begin{align}
-\dot{E}     &=  \frac{4 \pi (c/H)^{2}  c^{3} }{4 \pi G }  \left (  -  \dot{H}    +  h_{\textrm{B}}(t)  \right )    \notag \\
                &=  \frac{ c^{5} }{ G }  \left (  -  \frac{ \dot{H} } {H^{2}}    +  \frac{ h_{\textrm{B}}(t) }{H^{2}}  \right )    .
\label{eq:-dEdt3_A}
\end{align}
The two equations represent an energy flow across the Hubble horizon of a flat FRW universe.
Mimoso and Pav\'{o}n \cite{Mimoso2018} have discussed an equivalent equation in CCDM models in a non-flat FRW universe using an apparent horizon.

\section{Solutions for the present model with a power-law term} 
\label{Solution1}

In this appendix, the general solution for the present model with $H^{\alpha}$ terms is derived using a method used in Ref.\ \cite{Koma10}. 
The solution method is partially based on Refs.\ \cite{Koma45,Koma6}. 
We first consider $\alpha \neq 2$ and discuss $\alpha = 2$ later, where $\alpha$ is a real number.

The differential equation for the present model given by Eq.\ (\ref{eq:Back_power}) can be written as 
\begin{equation}
    \dot{H} = - \frac{3(1+w)}{2} H^{2}  \left ( 1 -  \frac{ f_{\alpha} (H)}{H^{2}}  \right )     .  
\label{eq:Back2_A}
\end{equation}
From Eq.\ (\ref{eq:Back2_A}), we have $(dH/da) a$ given by 
\begin{align}
\left ( \frac{dH}{da} \right )  a     &=        \frac{dH}{dt}    \frac{dt}{da} a  
                                                   = - \frac{3(1+w)}{2} H^{2}  \left ( 1 -  \frac{ f_{\alpha} (H) }{H^{2}}  \right )    \frac{a}{\dot{a}}                      \notag \\
                                                 &= - \frac{3(1+w)}{2} H  \left ( 1 -  \frac{ f_{\alpha} (H) }{H^{2}}  \right )                                                      .       
\label{eq:Back23_A}
\end{align}
We consider a matter-dominated universe, i.e., $w =0$, although $w$ is retained for generality.

In the present model, the extra driving term $ f_{\alpha} (H) $ given by Eq.\ (\ref{eq:fa2}) is 
\begin{equation}
        f_{\alpha} (H)  =   \Psi_{\alpha} H_{0}^{2} \left (  \frac{H}{H_{0}} \right )^{\alpha}  , 
\label{eq:fa2_A}
\end{equation}
where $\Psi_{\alpha}$ is a free parameter. 
Substituting Eq.\ (\ref{eq:fa2_A}) into Eq.\ (\ref{eq:Back23_A}) yields
\begin{align}
 \left ( \frac{dH}{da} \right )  a  &=  - \frac{3 (1+w)}{2} H  \left ( 1 -  \frac{ \Psi_{\alpha} H_{0}^{2} \left (  \frac{H}{H_{0}} \right )^{\alpha} }{H^{2}}  \right )                   \notag \\ 
                                               &=  - \frac{3 (1+w)}{2} H  \left ( 1 -   \Psi_{\alpha}  \left (  \frac{H}{H_{0}} \right )^{\alpha-2}   \right )                                  .                                 
\label{eq:Back24_A}
\end{align}
Now, the normalized Hubble parameter $\tilde{H}$ is defined as
\begin{equation}
 \tilde{H} \equiv \frac{H}{H_{0}}  .
\label{def_H_H0}
\end{equation}
Similarly, the normalized scale factor $\tilde{a}$ is defined as
\begin{equation}
 \tilde{a} \equiv \frac{a}{a_{0}} . 
\label{def_a_a0}
\end{equation}
Substituting  $ H = \tilde{H} H_{0} $ and $a = \tilde{a} a_{0} $ into Eq.\ (\ref{eq:Back24_A}) and arranging the resultant equation yields
\begin{equation}
 \left ( \frac{d \tilde{H}  }{d \tilde{a}  } \right )  \tilde{a}     =  - \frac{3 (1+w)}{2} \tilde{H}  \left ( 1 -   \Psi_{\alpha}  \tilde{H}^{\alpha-2}    \right )             .
\label{eq:Back24_A_2}
\end{equation}
In addition, a parameter $N$ is defined by 
\begin{equation}
   N  \equiv \ln \tilde{a}, \quad \textrm{and therefore,}  \quad   dN   = \frac{ d \tilde{a} }{ \tilde{a} }  .
\label{eq:N_A}
\end{equation}
Note that the $N$ defined here is not the number of degrees of freedom.
Using Eq.\ (\ref{eq:N_A}), Eq.\ (\ref{eq:Back24_A_2}) can be written as 
\begin{equation}
    \left ( \frac{d \tilde{H}  }{d N } \right )      =   - \frac{3 (1+w)}{2} \tilde{H}  \left ( 1 -   \Psi_{\alpha}  \tilde{H}^{\alpha-2}    \right )         .
\label{eq:Back24_A_N}
\end{equation}
When $w$, $\alpha$, and $\Psi_{\alpha}$ are constant, Eq.\ (\ref{eq:Back24_A_N}) can be integrated as 
\begin{equation}
    \int \frac{d \tilde{H}  }{  \tilde{H}  \left ( 1 -   \Psi_{\alpha}  \tilde{H}^{\alpha-2}    \right )      }       =   -   \frac{3 (1+w)}{2} \int dN         .        
\label{f_N_I}
\end{equation}
The solution is given by
\begin{align}
\ln  \left ( \frac{  \tilde{H}^{\alpha}  }{  \tilde{H}^{2} - \Psi_{\alpha} \tilde{H}^{\alpha}  } \right )^{\frac{1}{\alpha -2}}   
&=   -  \frac{3 (1+w)}{2}  N  + C                                                                                                                                             \notag \\
&= -  \frac{3 (1+w)}{2} \ln \tilde{a} + C        , 
\label{f_N_Solve}
\end{align}
where $C$ is an integral constant.
From Eqs.\ (\ref{def_H_H0}) and (\ref{def_a_a0}), the present values of $\tilde{H}$ and $\tilde{a}$ are $1$.
Substituting $\tilde{H} =1$ and $\tilde{a} =1$ into Eq.\ (\ref{f_N_Solve}), the integral constant $C$ can be written as 
\begin{equation}
 C =    \ln  \left ( \frac{  1}{  1 - \Psi_{\alpha} } \right )^{\frac{1}{\alpha -2}}    . 
\label{IntC}
\end{equation}
Substituting Eq.\ (\ref{IntC}) into Eq.\ (\ref{f_N_Solve}) and solving the resultant equation with respect to $\tilde{a} \equiv a/a_{0}$ yields
\begin{equation}
       \frac{a}{a_{0}} =  \left [ \frac{  1- \Psi_{\alpha}  }{  (H/H_{0})^{2-\alpha} - \Psi_{\alpha}  } \right ]^{\frac{2}{3 (1+w) (2-\alpha)}}   , 
\label{eq:Sol_aa0_power}
\end{equation}
and solving this equation with respect to $\tilde{H} \equiv H/H_{0}$ yields
\begin{equation}
    \left ( \frac{H}{H_{0}} \right )^{2-\alpha}  =   (1- \Psi_{\alpha})   \left ( \frac{a}{a_{0}} \right )^{ - \frac{3 (1+w) (2-\alpha)}{2}  }  + \Psi_{\alpha}      .
\label{eq:Sol_HH0_power2}
\end{equation}
When $\alpha \neq 2$, Eqs.\ (\ref{eq:Sol_aa0_power}) and (\ref{eq:Sol_HH0_power2}) are the general solutions for the present model.
Equation\ (\ref{eq:Sol_aa0_power}) is briefly described in a previous study \cite{Koma11}.

When $\alpha = 2$, Eq.\ (\ref{eq:Back24_A_N}) is written as 
\begin{equation}
    \left ( \frac{d \tilde{H}  }{d N } \right )      =   - \frac{3 (1+w)}{2}   ( 1 -   \Psi_{\alpha}  )   \tilde{H}       .
\label{eq:Back24_A_N_H2}
\end{equation}
This equation is integrated as 
\begin{equation}
    \int \frac{d \tilde{H}  }{  \tilde{H}  }       =   -   \frac{3 (1+w)  ( 1 -   \Psi_{\alpha}  )  }{2} \int dN         .        
\label{f_N_I}
\end{equation}
The solution is 
\begin{align}
\ln   \tilde{H}   &= - \frac{3 (1+w) ( 1 -   \Psi_{\alpha}  )  }{2}  N  + C_{2}     \notag \\
                      &= - \frac{3 (1+w) ( 1 -   \Psi_{\alpha}  )  }{2} \ln \tilde{a} + C_{2}        , 
\label{f_N_Solve_H2_2}
\end{align}
where $C_{2}$ is an integral constant.
Substituting $\tilde{H} =1$ and $\tilde{a} =1$ into Eq.\ (\ref{f_N_Solve_H2_2}) yields $C_{2}=0$.
From Eq.\ (\ref{f_N_Solve_H2_2}) and $C_{2}=0$, the solution for $\alpha =2$ is written as
\begin{equation}
     \frac{H}{H_{0}}  =     \left ( \frac{a}{a_{0}} \right )^{ - \frac{3 (1+w) (1- \Psi_{\alpha}) }{2}  }    , 
\label{eq:Sol_HH0_aa0_H2_A}
\end{equation}
where $\tilde{H} \equiv H/H_{0}$ and $\tilde{a} \equiv a/a_{0}$ are used.
The background evolution of the universe for $\alpha = 2$ has been discussed in, e.g., Ref.\ \cite{Koma45}.

\section{Calculation of $3 \dot{H}^{2}   - \ddot{H} H $ for the present model}
\label{Calculation2}

As discussed in Sec.\ \ref{dS2dt2}, $\ddot{S}_{\rm{BH}}$ includes $3 \dot{H}^{2}   - \ddot{H} H $ terms.
We calculate $3 \dot{H}^{2}   - \ddot{H} H $ for the present model.
To this end, a normalized parameter $B_{1}$ is defined by
\begin{equation}   
     B_{1} \equiv \frac{ 3 \dot{H}^{2}   - \ddot{H} H}{ H_{0}^{4} } =  3  \left ( \frac{\dot{H}}{ H_{0}^{2} } \right )^{2}   -   \frac{H}{ H_{0}} \left ( \frac{ \ddot{H} }{ H_{0}^{3} } \right )      .
\label{eq:B1set_A}
\end{equation}
To calculate the right-hand side of this equation, we first calculate $\ddot{H}$ from Eq.\ (\ref{eq:Back_power_12}), which is satisfied for all $\alpha$. 
Differentiating Eq.\ (\ref{eq:Back_power_12}) with respect to $t$ yields
\begin{align}  
    \ddot{H} &= \frac{d}{dt} \dot{H} = \frac{d}{dt} \left ( - \frac{3}{2} H^{2}  +  \frac{3}{2}  \Psi_{\alpha} H_{0}^{2}  \left (  \frac{  H }{  H_{0} } \right )^{\alpha}      \right )       \notag \\
                 &= - 3H \dot{H} +  \frac{3}{2}  \Psi_{\alpha}   H_{0}^{2}  \alpha   \left (  \frac{H}{H_{0}} \right )^{\alpha -1}  \frac{ \dot{H} }{H_{0}}      \notag \\
                 &= - 3 \dot{H} H_{0}   \left (   \frac{H}{H_{0}}  -  \frac{1}{2}  \Psi_{\alpha}    \alpha   \left (  \frac{H}{H_{0}} \right )^{\alpha -1}   \right )     .
\label{eq:d2H_power_1_A}
\end{align}
Substituting Eq.\ (\ref{eq:d2H_power_1_A}) into Eq.\ (\ref{eq:B1set_A}) yields
\begin{align}  
  B_{1} &=  3  \left ( \frac{\dot{H}}{ H_{0}^{2} } \right )^{2}   -   \frac{H}{ H_{0}}   \left [  -  \frac{ 3 \dot{H}}{ H_{0}^{2} }    \left (   \frac{H}{H_{0}}  -  \frac{ \Psi_{\alpha} \alpha }{2}    \left (  \frac{H}{H_{0}} \right )^{\alpha -1}   \right )     \right ]    \notag\\
     &=  3   \frac{\dot{H}}{ H_{0}^{2} } \left [  \frac{\dot{H}}{ H_{0}^{2} }  +  \left (   \frac{H}{ H_{0}} \right )^{2}   -  \frac{ \Psi_{\alpha} \alpha }{2}    \left (  \frac{H}{H_{0}} \right )^{\alpha}  \right ]   \notag\\
     &=  3   \left ( \frac{\dot{H}}{ H_{0}^{2} } \right )  B_{2}    ,
\label{eq:B1set2_A}
\end{align}
where $B_{2}$ is  
\begin{align}  
  B_{2} &=   \frac{\dot{H}}{ H_{0}^{2} }  +  \left (   \frac{H}{ H_{0}} \right )^{2}   -  \frac{ \Psi_{\alpha} \alpha }{2}    \left (  \frac{H}{H_{0}} \right )^{\alpha}    .
\label{eq:B2set_A}
\end{align}
Substituting Eq.\ (\ref{eq:Back_power_12}) into Eq.\ (\ref{eq:B2set_A}), $B_{2}$ is given by
\begin{align}  
  B_{2}  &=    \frac{  - \frac{3}{2} H^{2}  +  \frac{3}{2}  \Psi_{\alpha} H_{0}^{2}  \left (  \frac{  H }{  H_{0} } \right )^{\alpha}         }{ H_{0}^{2} }  +  \left (   \frac{H}{ H_{0}} \right )^{2}   -  \frac{ \Psi_{\alpha} \alpha }{2}    \left (  \frac{H}{H_{0}} \right )^{\alpha}     \notag\\
           &= - \frac{1}{2} \left (  \frac{  H }{  H_{0} } \right )^{2}     +  \frac{\Psi_{\alpha}  (3 - \alpha )}{2}    \left (  \frac{  H }{  H_{0} } \right )^{\alpha}    \notag\\
           &=   \frac{1}{2}  \left (  \frac{  H }{  H_{0} } \right )^{2}   \left [ -  1 +  \Psi_{\alpha}  (3 - \alpha )    \left (  \frac{  H }{  H_{0} } \right )^{\alpha -2}   \right ]  .
\label{eq:B2set2_A}
\end{align}
In addition, substituting Eq.\ (\ref{eq:B2set2_A}) into Eq.\ (\ref{eq:B1set2_A}), $B_{1}$ is written as
\begin{align}  
  B_{1}   &=  3   \left ( \frac{\dot{H}}{ H_{0}^{2} } \right )  \frac{1}{2}  \left (  \frac{  H }{  H_{0} } \right )^{2}   \left [ -  1 +  \Psi_{\alpha}  (3 - \alpha )    \left (  \frac{  H }{  H_{0} } \right )^{\alpha-2}   \right ]   \notag\\  
            &=  \frac{3}{2}    \left ( \frac{ - \dot{H}}{ H_{0}^{2} } \right )   \left (  \frac{  H }{  H_{0} } \right )^{2}   \left [  1 -  \Psi_{\alpha}  (3 - \alpha )    \left (  \frac{  H }{  H_{0} } \right )^{\alpha-2}   \right ]   .
\label{eq:B1set3_A}
\end{align}
From Eqs.\ (\ref{eq:B1set_A}) and (\ref{eq:B1set3_A}), we have
\begin{align}  
   3 \dot{H}^{2}   - \ddot{H} H     &=  \frac{3}{2}     ( - \dot{H} ) H^{2}   \left [ 1 -  \Psi_{\alpha}  (3 - \alpha )    \left (  \frac{  H }{  H_{0} } \right )^{\alpha-2}   \right ]   .
\label{eq:B1set5_A}
\end{align}
Applying this equation, $\ddot{S}_{\rm{BH}}$ is calculated from Eq.\ (\ref{eq:d2SB_1_11}), as examined in Sec.\ \ref{dS2dt2}.
Equation\ (\ref{eq:B1set5_A}) is satisfied for all $\alpha$ in the present model.

\end{document}